\documentclass[a4paper,fleqn]{cas-dc}

\usepackage[numbers]{natbib}

\usepackage{graphicx,subfig,epsfig}
\usepackage{amssymb,amsmath,amsbsy,bm,bbm}
\usepackage{algorithm,algpseudocode,theorem}
\usepackage{natbib}

\newtheorem{theorem}{Theorem}
 \newproof{pf}{Proof}
 
\def\tsc#1{\csdef{#1}{\textsc{\lowercase{#1}}\xspace}}
\tsc{WGM}
\tsc{QE}
\tsc{EP}
\tsc{PMS}
\tsc{BEC}
\tsc{DE}

\begin{document}
\let\WriteBookmarks\relax
\def\floatpagepagefraction{1}
\def\textpagefraction{.001}
\shorttitle{Dose-dependent treatment of engineered T cell therapy in cancer}
\shortauthors{H Cho, Z Wang, D Levy}

\title [mode = title]{Study of dose-dependent combination immunotherapy using engineered T cells and IL-2 in cervical cancer}                      



\author[1]{Heyrim Cho}
\ead{heyrimc@ucr.edu}


\address[1]{Department of Mathematics, University of California, Riverside, CA 92521}

\author[2]{Zuping Wang}

\ead{zpwang@umd.edu}


\address[2]{Department of Mathematics, University of Maryland, College Park, College Park, MD 20742}

\author%
[2,3]{Doron Levy}
\cormark[1]
\fnmark[1]
\ead{dlevy@umd.edu}

\address[3]{Center for Scientific Computation and Mathematical Modeling (CSCAMM), University of Maryland, College Park, College Park, MD 20742}

\cortext[cor1]{Corresponding author}
\fntext[fn1]{The work of DL was supported in part by the National Science Foundation under Grant Number DMS-1713109 and by the Jayne Koskinas Ted Giovanis Foundation.}


\begin{abstract}
Adoptive T cell based immunotherapy is gaining significant traction in cancer treatment. Despite its limited success, so far, in treating solid cancers, it is increasingly successful, demonstrating to have a broader therapeutic potential. 
In this paper we develop a mathematical model to study the efficacy of engineered T cell receptor (TCR) T cell therapy targeting the E7 antigen in cervical cancer cell lines. We consider a dynamical system that follows the population of cancer cells, TCR T cells, and IL-2.  We demonstrate that there exists a TCR T cell dosage window for a successful cancer elimination that can be expressed in terms of the initial tumor size. We obtain the TCR T cell dose for two cervical cancer cell lines: 4050 and CaSki. 
Finally, a combination therapy of TCR T cell and IL-2 treatment is studied. We show that certain treatment protocols can improve therapy responses in the 4050 cell line, but not in the CaSki cell line. 
\end{abstract}




\maketitle

\section{Introduction}

Adoptive T cell therapy, also called cellular adoptive immunotherapy or T cell transfer therapy, is an immunotherapy that uses T cells to help patients overcome diseases such as cancer. In adoptive T cell therapy, T cells are typically collected from the patient, engineered to improve their ability to target the patient's cancer cells, and cultured to large numbers before being introduced back to the patient \citep{Paucek2019,Hinrichs2016}. Adoptive T cell therapy includes tumor-infiltrating lymphocyte (TIL) therapy \citep{Dudley2003,Rohaan2018}, T cell receptor (TCR) T cell therapy \citep{Govers2010,Jin2018,Zhang2019}, and chimeric antigen receptor (CAR) T cell therapy \citep{June2018,Benmebarek2019}. The use of immune cells from donors is being studied as well. This therapy has been of growing interest as a potential anti-cancer treatment in recent years. However, at present, its applicability has been mostly limited to blood cancers. Recent studies are focusing on broadening the applicability of the therapy to other types of cancer including solid tumors \citep{June2018,Paucek2019}. 
Other issues that are being investigated are the enhancement of the  T cell production and activation, including the selection of T cell subsets, as well as adjusting the clinical protocols. 

Another challenge of T cell therapies include a phenomena known as "exhaustion" \citep{DWherry2011, Kurachi2019,Lynn2019}. 
Although Tumor-infiltrating CD8+ T cells can attack tumors, high and sustained antigen exposure often leads CD8+ T cells to a gradual loss of their functionality. Exhausted tumor-infiltrating lymphocytes (TILs) are characterized by progressive and hierarchical loss of effector functions such as defects in production of IL-2, IFN-$\gamma$ and chemokines, high proliferative capacity and \emph{ex vivo} killing, sustained upregulation and co-expression of multiple inhibitory receptors including the cytotoxic T lymphocyte-associated protein 4 (CTLA-4) and programmed cell death protein 1 (PD-1).
They are also characterized by altered expression and use of key transcription factors, metabolic dysregulation, and by their inability to transition to quiescence and acquire antigen-independent memory T cell homeostatic responsiveness \citep{DWherry2011,Kurachi2019,Wherry2015,McLane2019}. Whereas blocking CTLA-4 and PD-1 inhibitory receptor can reinvigorate exhausted T cell responses, resulting in improved proliferation and function \citep{Barber2006,Sharma2015}; deletion of nuclear factor TOX, TOX2 and members of the NR4A family in  tumor-specific T cells in tumors can abrogate the exhaustion program \citep{Scott2019,Seo2019}. These studies demonstrated that T cell exhaustion was reversible rather than a terminal state. Thus, a better understanding of mechanisms of T cell development could potentially be used to engineer T cells and control T cell exhaustion.
 
Mathematical models that describe the interaction of cancer and immune cells date back to \cite{Kuznetsov1994}, where a dynamical system involving the tumor and cytotoxic T lymphocytes was studied. Periodic treatment and time delay were included to model persistant oscillations in \citep{Costa2003}, followed by a stability analysis in \citep{DOnofrio2008}. Further developments of the model included adding new types of cells, such as Natural Killer (NK) cells and normal cells, as well as various cytokines \citep{Kirschner1998,DePillis2003,Moore2004}. These models capture the immune escape of tumors and explain multiple equilibrium phases of coexisting immune cells and cancer cells. Although the parameterization and analysis become difficult, dynamical systems in higher dimensions, stochastic models, agent-based and cellular automata models, as well as partial differential equations have all been used to test different biological hypotheses including multiple immune cell populations and signaling molecules \citep{DePillis2007,DePillis2009,PKim2008,Piccoli2007}. The recent surge of clinical trials and the success of adoptive immunotherapies inspired the adaptation of these  mathematical models to the new therapies \citep{Konstorum2017}, including adoptive T cell therapies \citep{Talkington2018}. 
For instance, CD19 CAR T cell therapy targeting acute lymphoblastic leukemia is modeled in \cite{Mostolizadeh2018} with a dynamical system that also includes healthy B cell populations and circulating lymphocytes. However, this model was not calibrated with experimental data.
CD19 CAR T cell therapy applied to chronic lymphocytic leukemia is studied in \cite{Hardiansyah2019} where the relationships between CAR T cell doses and disease burden are being explored. 
To study the cytokine release syndrome (one of the primary side effects of CAR T cell therapy) a dynamical system of nine cytokines responding to CAR T cell therapy is developed and studied in \cite{Hopkins2018}.
More recently, CAR T cell therapies for glioblastoma are modeled in \cite{Sahoo2019}. 
Another approach to immunotherapy, immune checkpoint inhibitor therapies are modeled in \citep{Peskov2019,Nikolopoulou2018,Radunskaya2018}. 

In this paper, we focus on engineered T cells therapy
targeting human papilloma virus (HPV) E7 antigen in solid tumor that is  developed and studied in \cite{Jin2018}. The viral oncoprotein E7 is an attractive therapeutic target due to its constructive expression in HPV-associated cancers but not in healthy tissues. Through a uterine cervix biopsy of a woman with cervical intraepithelial neoplasia II/III, Jin \emph{et al.} discovered an HPV-16 E7 antigen-specific, HLA-A*02:01-restricted TCR. By reversing $\alpha$ and $\beta$ chain and adding disulfide bond and hydrophobic substitutions, this modified TCR demonstrated high avidity for the target epitope and no perceptible cross-reactivity against human peptides. \emph{In vitro}, genetic engineered T cells that express E7-targeting TCR demonstrated effector T cell functions, including IFN-$\gamma$ production and CD8 coreceptor–independent tumor cell killing. \emph{In vivo}, immunodeficient, NOD/SCID $\gamma$ (NSG) mouse model, E7 TCR-transduced T cells  mediated regression of the CaSki cell line (a HPV-16$^+$ cervical cancer) at doses of $1 \times 10^6$ or $1 \times 10^7$ cells and repressed the 4050 cell line growth (a HPV-16$^+$ oropharyngeal cancer) at doses of $1 \times 10^7$ cells. This antitumor activity could be enhanced by the addition of systemic IL-2.

The goal of this study is to demonstrate the potential role of mathematical modeling in improving the administration of adoptive TCR T cell therapy for cancer treatment. In Section~\ref{sec:model}, we present a cancer-immune interaction model that follows the dynamics of cancer cells, TCR engineered T cells, and the cytokine IL-2. We describe the model parameters and assumptions, and the procedure of sequential model calibration. In Section~\ref{sec:stability}, stability analysis of the model is conducted, resulting with conditions for therapy success. In Section~\ref{sec:dose}, we study the dose-dependent response of two cancer cell lines, 4050 and CaSki to TCR T cell treatment.  We demonstrate the existence of a TCR T cell dose-dependent therapeutic window. The combination of TCR T cell and IL-2 treatment is studied in Section~\ref{sec:result3}, where we investigate the effect of different IL-2 treatment schedules, and show that IL-2 treatment given in a longer period of time is effective in the 4050 cell line, but not in the CaSki cell line. A summary and future outlook are provided in Section~\ref{sec:conclusion}.

\section{Model} 
\label{sec:model}
We denote cancer cells by $C(t)$, TCR engineered T cells by $T(t)$, and the cytokine IL-2 by $I(t)$.
The dynamics of cancer-immune interactions is then modeled as
\begin{align}
    \dot{C} & =  aC(1-bC) - nTC , \label{eq:ode1} \\
    \dot{T} & =  s_T(t)  - dT + pT\frac{C}{g+C} - mCT + p_1T\frac{I}{g_1+I} , \label{eq:ode2} \\  
	\dot{I}  & =  s_I(t)  - kI + p_2I\frac{T}{g_2+T}. \label{eq:ode3}
\end{align}

The system (\ref{eq:ode1})--(\ref{eq:ode3}) is adapted from existing models describing the interaction of cancer cells and T cells \citep{Kuznetsov1994, Kirschner1998, DePillis2006}.  

In Eq.~(\ref{eq:ode1}), the cancer is assumed to follow a logistic growth with growth rate $a$ and tumor capacity $1/b$. The interaction between cancer cells and T cells results with a tumor death that is induced by the T cells with death rate $n$.

The TCR T cell therapy is represented by a source term $s_T(t)$ in Eq.~(\ref{eq:ode2}). These cells die exponentially at rate $d$. 
The engineered TCR T cells are activated by the presence of the cancer cells with E7 antigen, that is modeled with the parameter $p$ denoting the rate of proliferation of T cells induced by cancer. 
 The saturation of this proliferation for large values of cancer cells follows a Michaelis-Menten dynamics, and is given by $g$, a parameter that represents the number of cancer cells that reduce the maximal T cell activation by half.
 In addition, we assume that the interaction between cancer and T cells further reduces the T cell population at a rate $m$.

In Eq.~(\ref{eq:ode3}), the IL-2 therapy is modeled similarly to Eq.~(\ref{eq:ode2}) with a source term $s_I(t)$ and a decay rate $k$. 
The model includes the interaction between IL-2 and T cells, where we assume that the two populations stimulate each other. 
Although the effect of IL-2 on T cells is known to be both stimulating and inhibitory \cite{Busse2010}, we assume that the net effect is positive.
This is supported by the data of  \cite{Jin2018}. The rates of T cell and IL-2 production stimulated by each other are denoted as $p_1$ and $p_2$, respectively. We also assume saturation in the growth dynamics of T cells and IL-2 with parameters $g_1$ and $g_2$. 

The treatments are given as follows. The T cell treatment is given once at the initial time $t_0=0$, while the IL-2 treatment is given $d$ times at times $t_1,...,t_d$.
Accordingly, the source terms are defined as 
$$s_T(t) =  \bar{s}_1 \mathbf{1}_{t=t_0}(t), \quad 
s_I(t) =  \sum_{i=1}^{d} \bar{s}_2 \mathbf{1}_{t=t_i}(t). $$

The model parameters and their biological interpretations are summarized in Table \ref{Tbl:param1}.

\begin{table} 
	\begin{tabular}{|c|l|c|c|c|c|} \hline 
	parameter & biological meaning \\ \hline 
	$ a $  &  tumor proliferation rate   \\ 
	$ b $  &  inverse of tumor carrying capacity   \\ 
	\hline \hline 
	$ n $  & tumor death rate induced by T cells  \\
	$ d $  & death rate of T cells  \\
	$ p $  & rate of T cell proliferation induced by tumor  \\ 
	$ g $  & steepness coefficient of  T cell recruitment  \\
	$ m $  & T cell inactivation rate induced 
	by tumor  \\ \hline\hline 
	$p_1 $ & rate of T cell proliferation stimulated by IL-2  \\
	$g_1 $ &  steepness of T cell proliferation curve by IL-2  \\
	$p_2 $ &  rate of IL-2 production by T cell and tumor 
	\\ & interaction   \\
	$g_2 $ & steepness of IL-2 production curve   \\
	$k$  &  natural decay rate of IL-2      \\  \hline 
	\end{tabular}
	\caption{Model parameters and their biological interpretation}
	\label{Tbl:param1}
\end{table}

\subsection{Sequential model calibration}

The experimental data in \cite{Jin2018} was obtained in three experimental settings: (1) cancer growth without treatment; (2) TCR T cell treatment; and (3) a combination of TCR T cell and IL-2 treatments. These experiments allow us to sequentially estimate the model parameters, and ensure their robust identification.
The ranges of parameters found in the literature are presented in Table~\ref{Tbl:param2} with references. 
We employ a Markov chain Monte Carlo (MCMC) algorithm, namely, delayed rejection adaptive metropolis (DRAM) \citep{Haario2006}. 
The fitted parameter values are shown in Table~\ref{Tbl:paramfit}.

\begin{table}
	\begin{tabular}{|c|c|c|c|c|c|} \hline 
	parameter & units & range \\ \hline 
	$ a $  & $day^{-1}$ & [0.01, 0.52]  \\ 
	$ b $  & $cell^{-1}$ & [$10^{-14}$, $10^{-4}$]   \\ 
	\hline \hline
	$ n $  & $day^{-1}cell^{-1} $ & [$3.4\cdot 10^{-10},3\cdot 10^{-7}$]   \\
	$ d $  & $day^{-1}$ & [0.01, 0.08]   \\
	$ p $  & $day ^{-1}$ & [0.1, 0.4]    \\ 
	$ g $  & $cell$  & $ 2.019 \cdot 10^{7}$   \\
	$ m $  & $day^{-1}cell^{-1}$ & [$10^{-12}, 5\cdot 10^{-7}$]    \\ \hline\hline 
	$p_1 $ &  $day^{-1} $& [0.124, 2.971]     \\
	$g_1 $ & $cells ^{2} $& $[2\cdot 10^{-6}, 2.5063\cdot 10^3]$      \\ 
	$p_2 $ &  $IU/cells^{-1} day^{-1}$ & [1, 5]   \\
	$g_2 $ & $ cells $ & $10^3$    \\
	$k$ & $day^{-1} $ &  [$5, 20$]   \\  \hline 
	\end{tabular}
	\caption{Model parameters and their ranges taken from
	\cite{Talkington2018, DePillis2006, DePillis2009, Rihan2014, Piotrowska2016, DePillis2007, Rihan2019}. 
	}
	\label{Tbl:param2}
\end{table}

	\begin{table}
		\begin{tabular}{|c|c|c|c|c|c|} \hline 
			parameters & Cancer 4050 & Cancer CaSki \\ \hline 
			$a$ & 0.1828   & 0.1212 \\
			$b$ & 2.6269e-7   & 1.5201e-7 \\ \hline \hline 
			$p$ & 0.1749 &  0.2144  \\
			$m$ & 7.2590e-8 & 3.3315e-8\\
			$n$  &  1.2883e-7 & 7.0924e-9 \\
			$d$  & 0.0212&  0.0330 \\
			$g$ &  1.7479e5 & 5.0880e5  \\ \hline \hline 
	        $p_1 $ & 0.21441 &  0.2040 \\
        	$g_1 $ & 1.6488  &  3718  \\
	        $p_2 $ & 0 & 0  \\
        	$g_2 $ & 1000 & 1000 \\
        	$k$    & 5 & 10 \\  \hline 		
        \end{tabular}
		\caption{Parameter values obtained with the MCMC algorithm using the \cite{Jin2018} data for the 4050 cell line and the CaSki cell line.}
		\label{Tbl:paramfit}
	\end{table}

	\begin{figure}
		\centerline{ 
			\includegraphics[width=4.5cm]{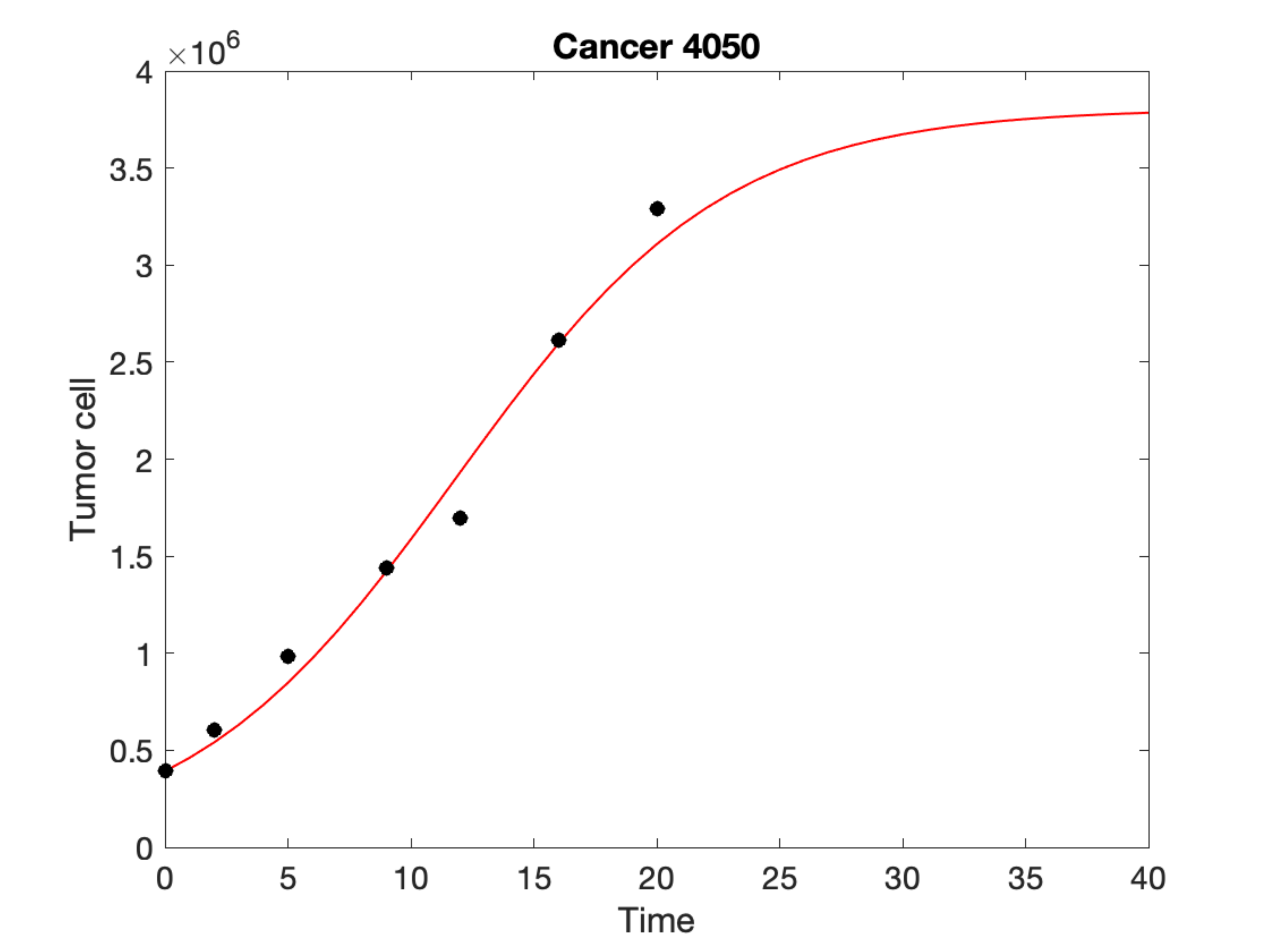}
			\includegraphics[width=4.5cm]{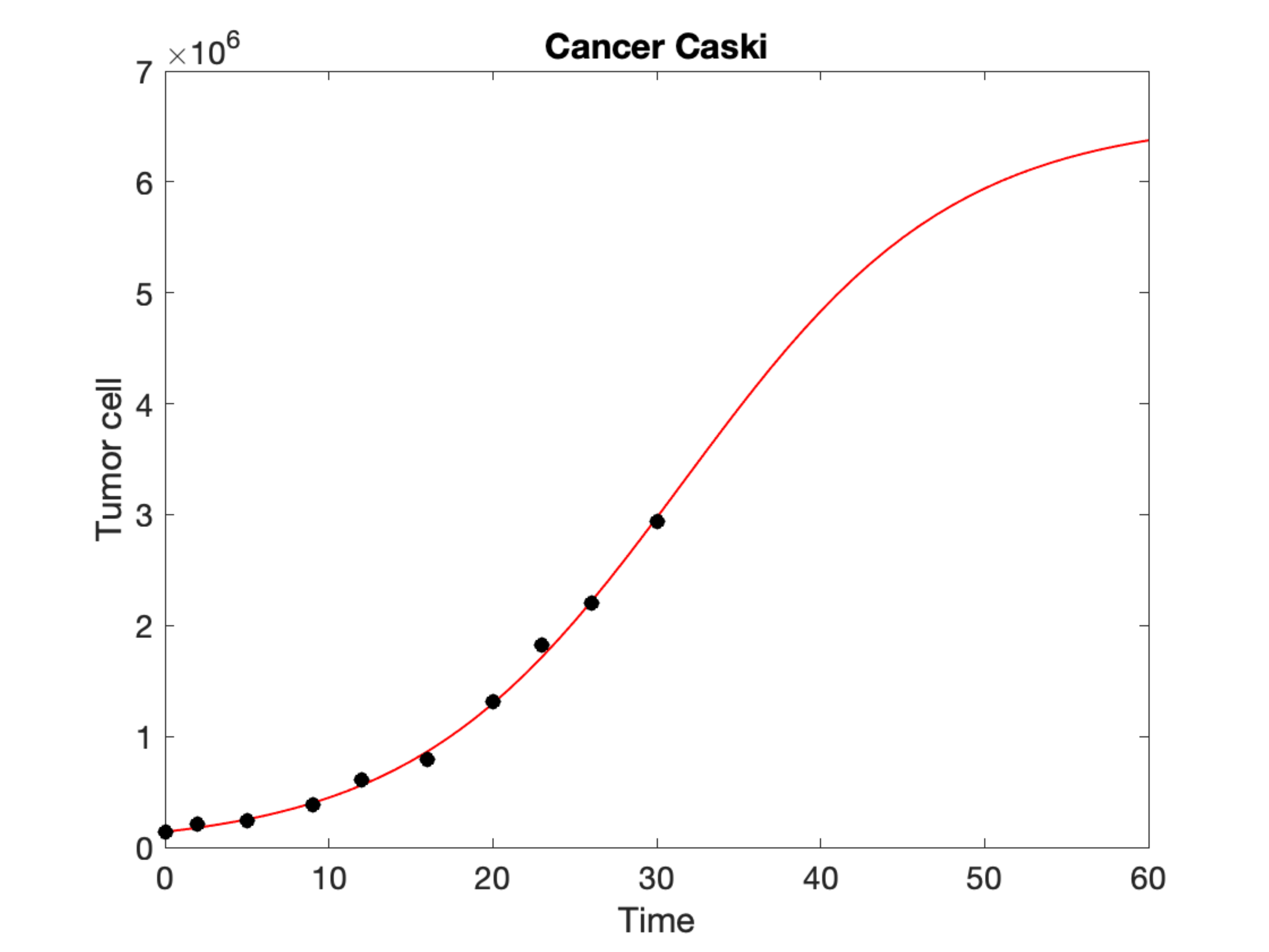}
			}
		\caption{A calibration of the tumor growth model Eq. \eqref{eq:ode1} to the  cancer growth data ($\bullet$) for the 4050 cell line (left) and the CaSki cell line (right) without treatment \cite{Jin2018}.
		} 
		\label{fig:data1} 
	\end{figure}

	\begin{figure}
		\centerline{ \footnotesize \rotatebox{90}{ \hspace{1cm} Cancer 4050}
			\includegraphics[width=8.5cm]{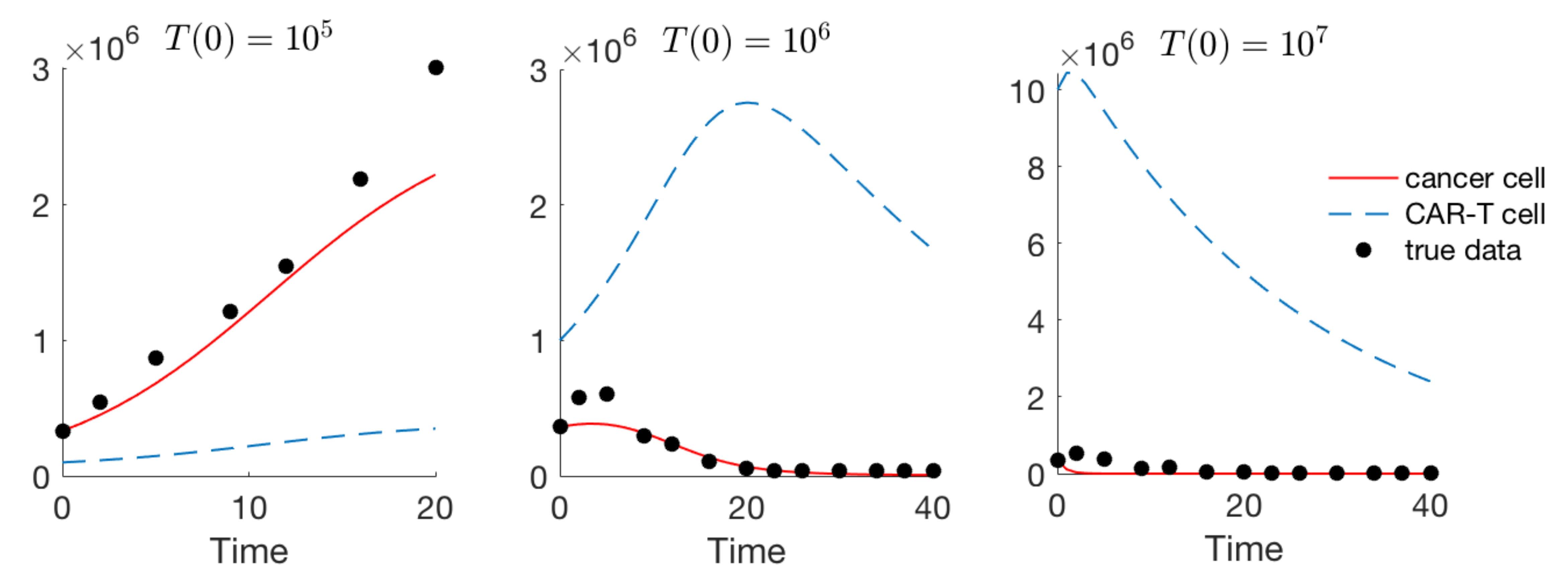}
		}
		\centerline{ \footnotesize \rotatebox{90}{ \hspace{1cm} Cancer CaSki}
			\includegraphics[width=8.5cm]{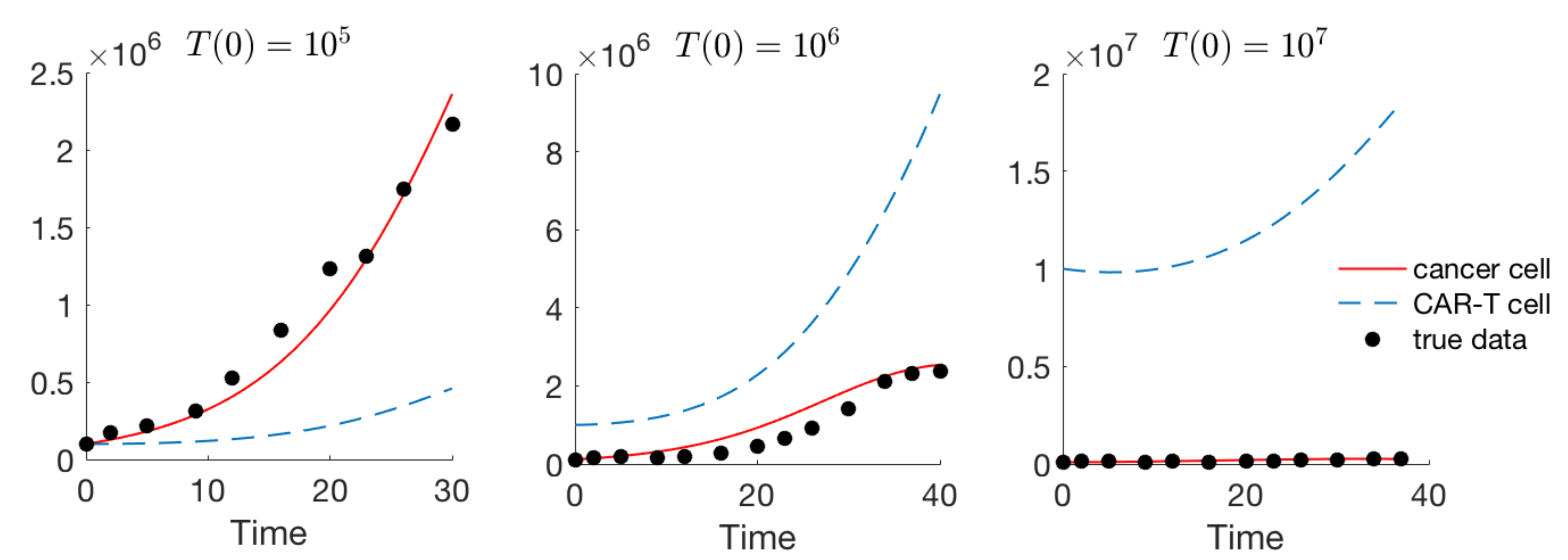}
		}		
		\caption{A calibration of the tumor--immune interaction model Eqs.~\eqref{eq:ode1}--\eqref{eq:ode2} to the cancer growth data with TCR treatment \cite{Jin2018}. Two cancer cell lines, 4050 (top) and CaSki (bottom), are treated with TCR T cell with dose $T(0)=10^5$, $10^6$, and $10^7$. 
		} 
		\label{fig:data2} 
	\end{figure}
	
Fig.~\ref{fig:data1} shows the experimentla data of tumor growth without treatment and the fitted logistic growth model~\eqref{eq:ode1} for the two cancer cell lines, 4050 and CaSki. The 4050 cell line reaches its full capacity around 30 days. This is faster compared to CaSki. The data with TCR engineered T cell treatment is shown in Fig. \ref{fig:data2}, where the dosage is given with three levels, $T(0)=10^5$, $10^6$, and $10^7$. 
In both 4050 and CaSki cell lines, the low dose of $T(0)=10^5$ does not prevent tumor progression. However, the higher dose of $T(0)=10^7$ results in tumor regression. The medium dose of $T(0)=10^6$ results with a tumor decay in the 4050 cell line despite its higher growth rate, but the tumor still grows in the CaSki cell line. This illustrates different susceptibilities depending on the type of cancer. 
The IL-2 treatment is shown to be effective in both cell lines, where the results are shown in section \ref{sec:result3}.

\section{Results} 
\subsection{Stability analysis reveals critical parameters for therapy success}
\label{sec:stability}

The experimental data reveals both scenarios of tumor progression and regression depending on the initial T cell dose. In this section, we study the steady states and their stability to gain a better understanding of the interaction between cancer and T cells in the model. 
We first focus on the steady states without the IL-2 treatment, that is, $(C,T,I) = (C,T,0)$. We assume that all the parameters are non-negative. We also assume that $s_T(t) = 0$ with a nonzero initial condition for the T cells, since the T cell treatment is given as an instant treatment at the initial time. 
The equilibrium states of the system satisfy
	\begin{align*}
 	 0  &= aC(1-bC) - nTC  = C[a(1-bC)-nT], \\
     0  &= - dT + pT\frac{C}{g+C} - mCT \\
      &= T\left[- d + p\frac{C}{g+C} - mC \right], 
	\end{align*}
where the linearized Jacobian is 
	$$L = \begin{pmatrix}
	a - 2abC-nT  & -nC \\
	\displaystyle{T\left[\frac{pg}{(g+C)^2} - m\right] } & \displaystyle{-d + p\frac{C}{g+C} - mC }
	\end{pmatrix}. $$
There exist four possible steady states $(T,\,C)$.
However, the steady states of interest are those with non-negative values. 
In particular, the equilibrium point $ (T,C) = (0,b^{-1}) $ is the case of tumor cells reaching their maximum capacity, while the T cells go extinct. This equilibrium state becomes stable 
when $-d + p(gb+1)^{-1} - mb^{-1} < 0$, which holds if 
\begin{equation}
p < \left(\frac{m}{b} +d\right)(gb+1).
\end{equation}
Otherwise it is unstable. 
This provides us with a necessary condition so that the T cell therapy is successful, that is, the minimum level of the proliferation rate of T cells that needs to be attained.

Another set of equilibrium points are $(C_i, T_i)$ for $i=1$ and $2$, where
$$C_i = \frac{(p-d-mg) \pm \sqrt {(p-d-mg)^2-4mgd}}{2m},$$ and  
$$T_i = \frac{a(1-bC_i)}{n}. $$
For these equilibrium points to be real and positive, it is required that $p-d-mg \geq 0$ and $(p-d-mg)^2-4mgd \geq 0$, or equivalently, 
\begin{equation}
(\sqrt{d} + \sqrt{mg})^2 \leq p.
\end{equation}
By ordering the points as $0 < C_1 < C_2$, 
we have $ T_1 > T_2 > 0 $. 
We denote $(C_1,\,T_1)$ as the T cell therapy success case that has a relatively smaller cancer size with a large T cell population. 
The conditions derived above classify the scenario of T cell therapy success, particularly relating the model parameters in terms of the cancer-induced proliferation rate $p$. 
In particular, T cell therapy always fails if the cancer induced proliferation rate is less than $(\sqrt{d} + \sqrt{mg})^2$. This is the minimum level of proliferation rate that should be achieved for the engineered T cells to be effective. 
On the other hand, if the T cell proliferation rate is larger than $(mb^{-1} +d)(gb+1)$, the tumor cannot achieve its maximum capacity and the therapy will result in a relatively small tumor equilibrium. 

\begin{theorem}
The T cell therapy fails regardless of the dose if $p < (\sqrt{d} + \sqrt{mg})^2$. 
The therapy succeeds if $ (\frac{m}{b} +d)(gb+1)< p$. 
If the T cell proliferation is in the range $ (\sqrt{d} + \sqrt{mg})^2 < p < (\frac{m}{b} +d)(gb+1)$, treatment success depends on the initial cancer size and T cell dosage. 
\end{theorem}
We note that this result can be used to restrict the search interval when estimating the model parameters. For instance, the experimental data in \cite{Jin2018} show both scenarios of T cell therapy success and failure, which indicates that the model should be able to capture both cases. 
Therefore, we should search for parameters that satisfy the condition
\begin{equation}
(\sqrt{d} + \sqrt{mg})^2 < p < \left(\frac{m}{b} +d\right)(gb+1).     
\label{eq:pCond}
\end{equation}
We remark that the trivial equilibrium state, $(T,\,C) = (0,\,0)$, and the relatively large tumor equilibrium, $(T_2, C_2)$, are both saddle points. 
The results are summarized in Table \ref{Tbl:paramRange} and the stability analysis and the proof of theorem 1 can be found in Appendix~\ref{sec:Apendix2}. 
 
\begin{table*}
		\begin{tabular}{|c|c|c|c|c|c|} \hline 
			condition & $(0,0)$ & $(0,1/b)$ & $(T_1, C_1)$ & $(T_2, C_2)$\\ \hline 
			  $p < (\sqrt{d} + \sqrt{mg})^2$  & saddle & stable & N/A & N/A \\
			$ (\sqrt{d} + \sqrt{mg})^2 < p < (\frac{m}{b} +d)(gb+1) $  & saddle  & stable & stable & saddle \\
			$ (\frac{m}{b} +d)(gb+1)< p  $  & saddle & unstable & stable & saddle \\ \hline
		\end{tabular}
		\caption{Stability of the equilibrium points as a function of the range of  the TCR T cell proliferation rate $p$.  }
		\label{Tbl:paramRange}
\end{table*}

\subsection{A study of the TCR T cell dose depending on the initial tumor size }
\label{sec:dose}

\begin{figure}
	\centerline{ 
		\includegraphics[width=4.5cm]{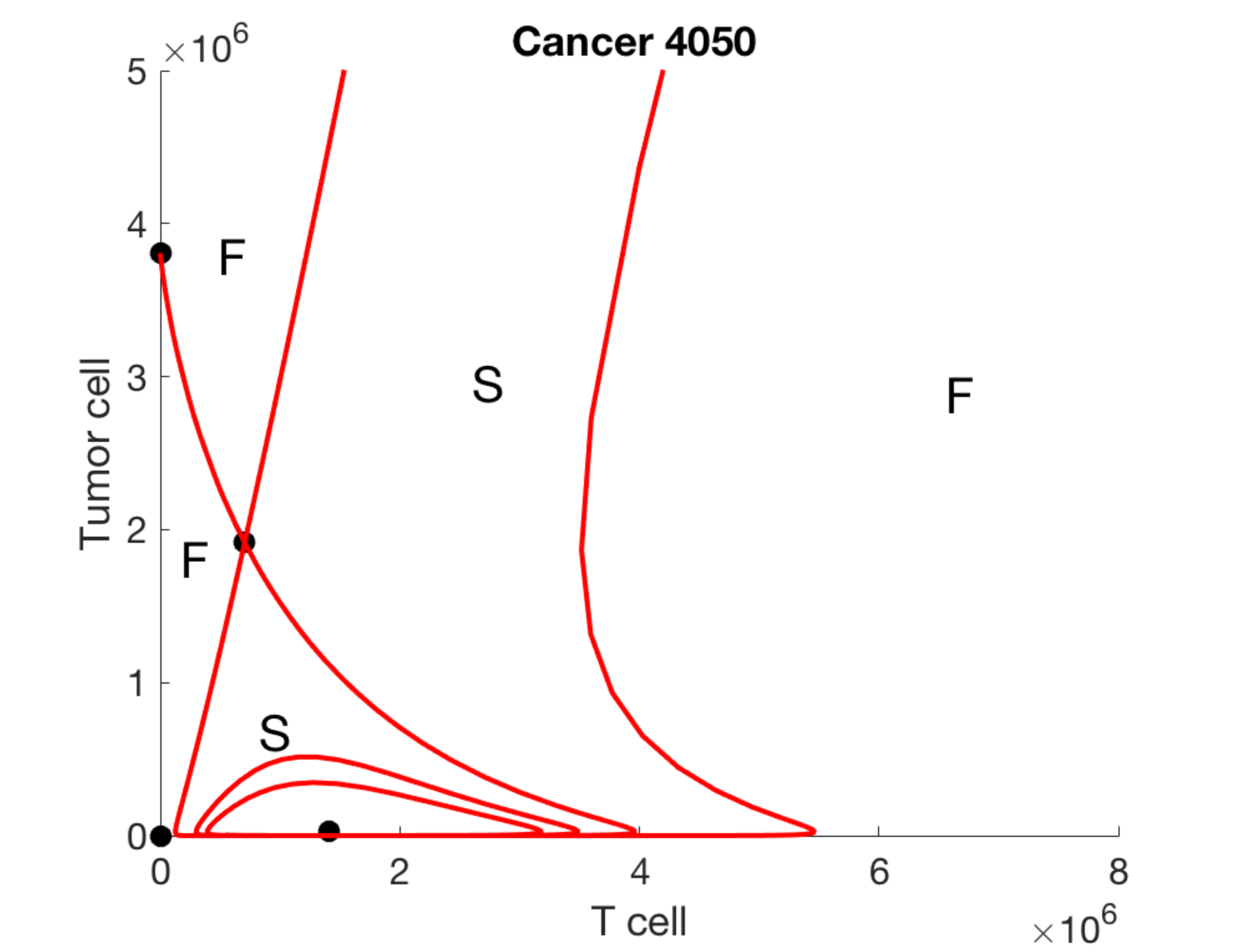}
		\includegraphics[width=4.5cm]{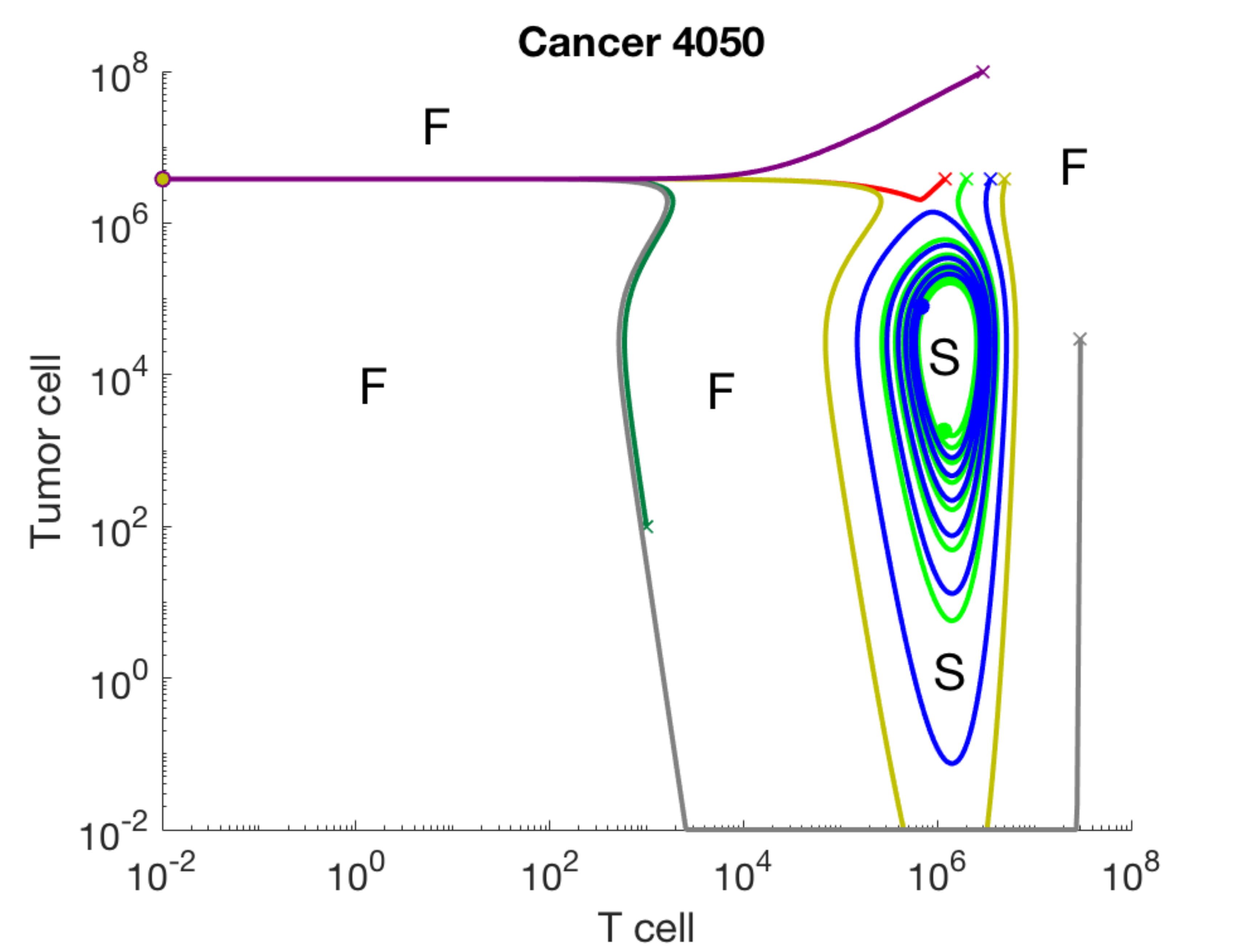}
	}
	\centerline{ 
		\includegraphics[width=4.5cm]{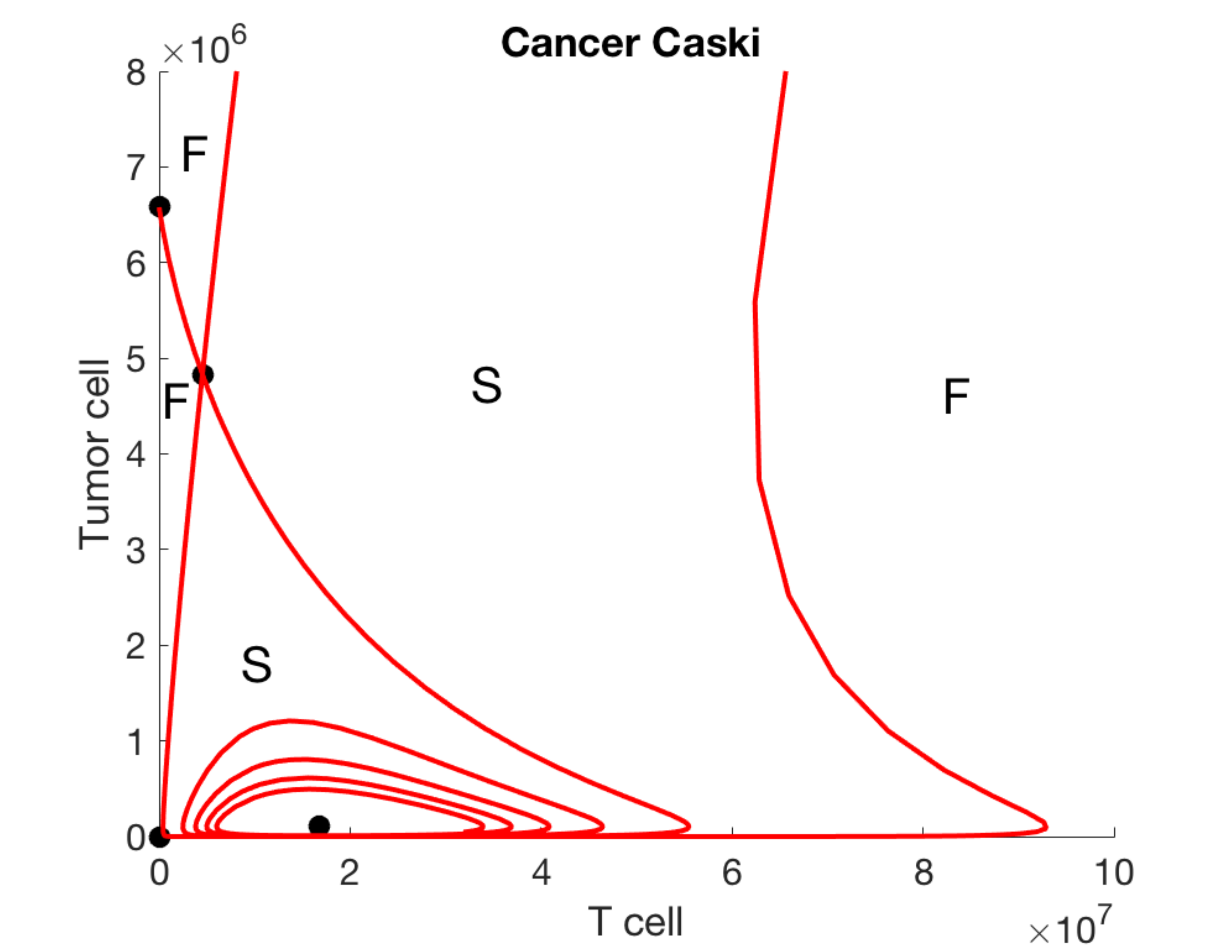}
		\includegraphics[width=4.5cm]{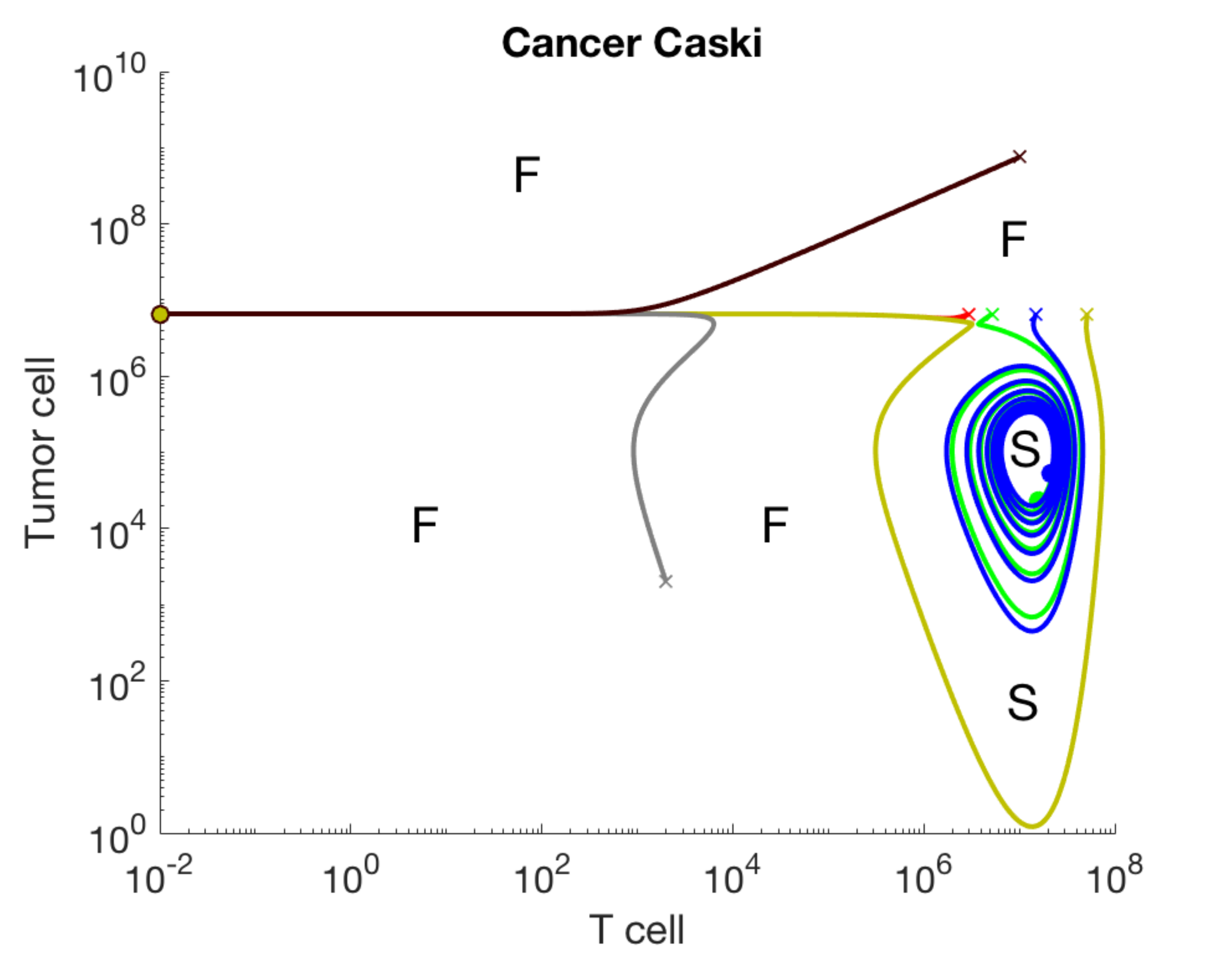}
	}
	\caption{The phase plane of the model~\eqref{eq:ode1}--\eqref{eq:ode2} for the 4050 cell line (top) and the CaSki cell line (bottom) in linear (left) and log (right) scale. 'F' denotes the initial tumor size and T cell dosage for which therapy fails and the tumor grows to its carrying capacity. 'S' denotes the case when T cell therapy is successful, and the tumor shrinks to $2.67 \times 10^4$ (4050 cell line) and $1.04 \times 10^5 $  (CaSki cell line). 
	} 
	\label{fig:phase} 
\end{figure}

\begin{figure}
\centerline{ 
		\includegraphics[width=4.5cm]{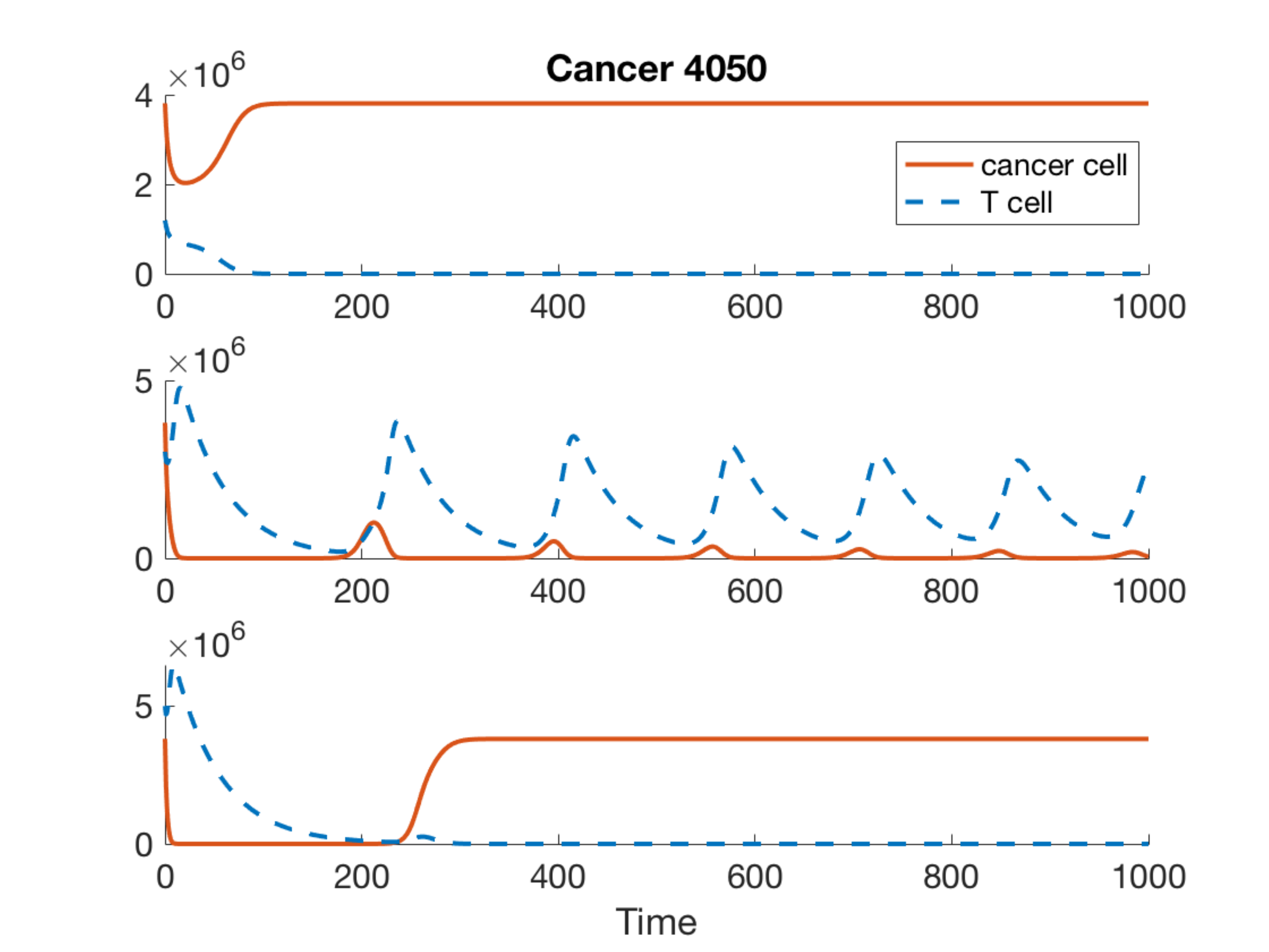}
		\includegraphics[width=4.5cm]{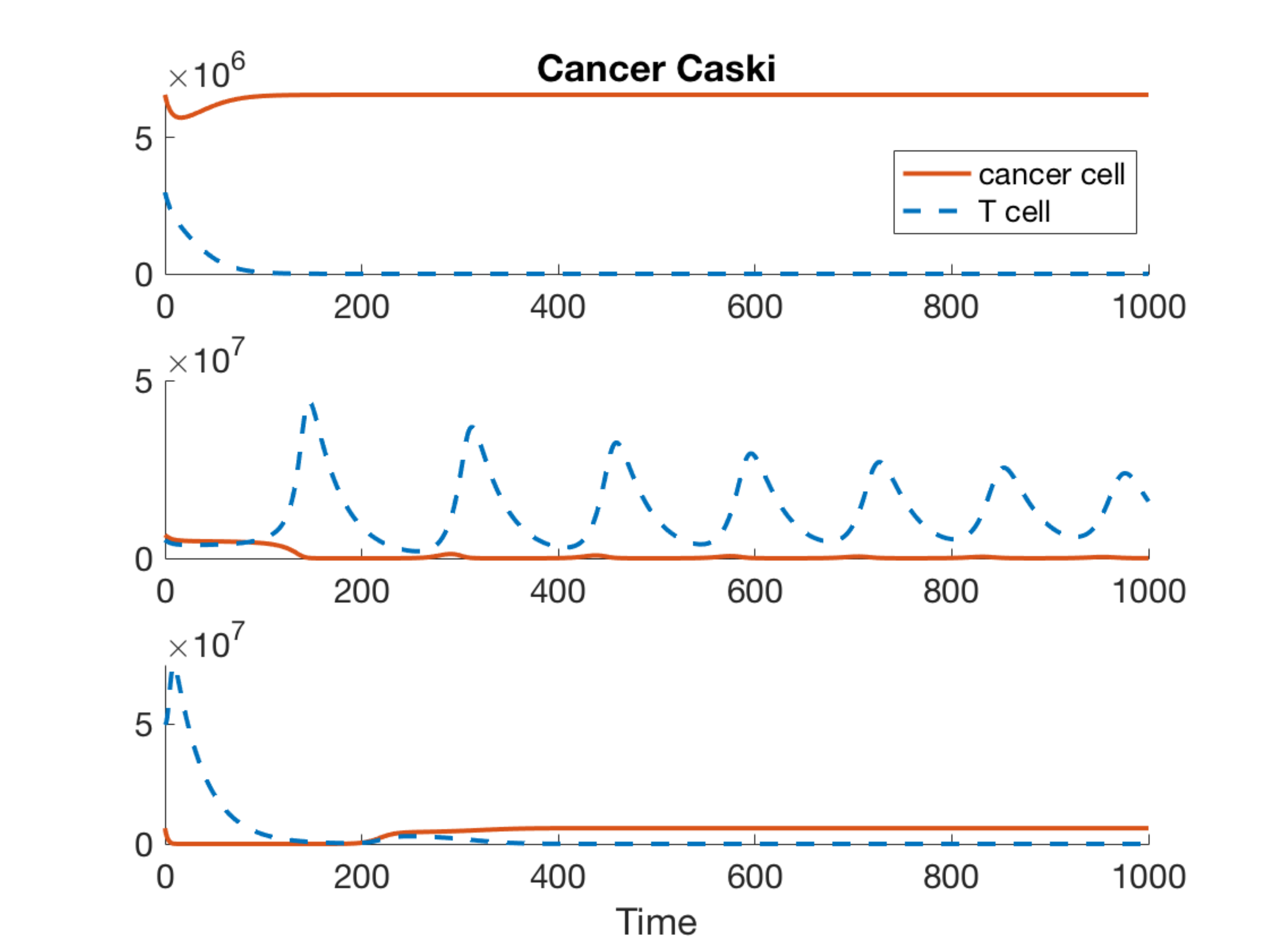}
	}
\centerline{ 
		\includegraphics[width=3cm]{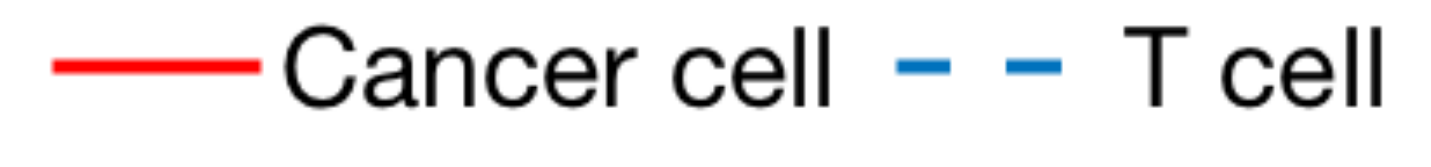}
		}
		\caption{The long-term dynamics of the 4050 cell line and the CaSki cell line with TCR T cell treatment for different dosages. The dosages are chosen to be below (top), within (middle), and above (bottom) the therapeutic window, that is, $( 1.3 \times 10^6, 3.8 \times 10^6)$ and $( 5.2 \times 10^6, 5.0 \times 10^7)$ for the cell lines 4050 and CaSki, respectively. 
	The initial tumor size is taken as its carrying capacity, and we show that even in its largest size, cancer can still be controlled by effective immune intervention and an appropriate dosage. 
	However, the T cells fail at low dosages, but also at very high dosage level, due to presumed premature T cell exhaustion and loss of anti-tumor activity.
	} 
	\label{fig:CnT_long} 
\end{figure}

\begin{figure}
\centerline{ 
		\includegraphics[width=4.2cm]{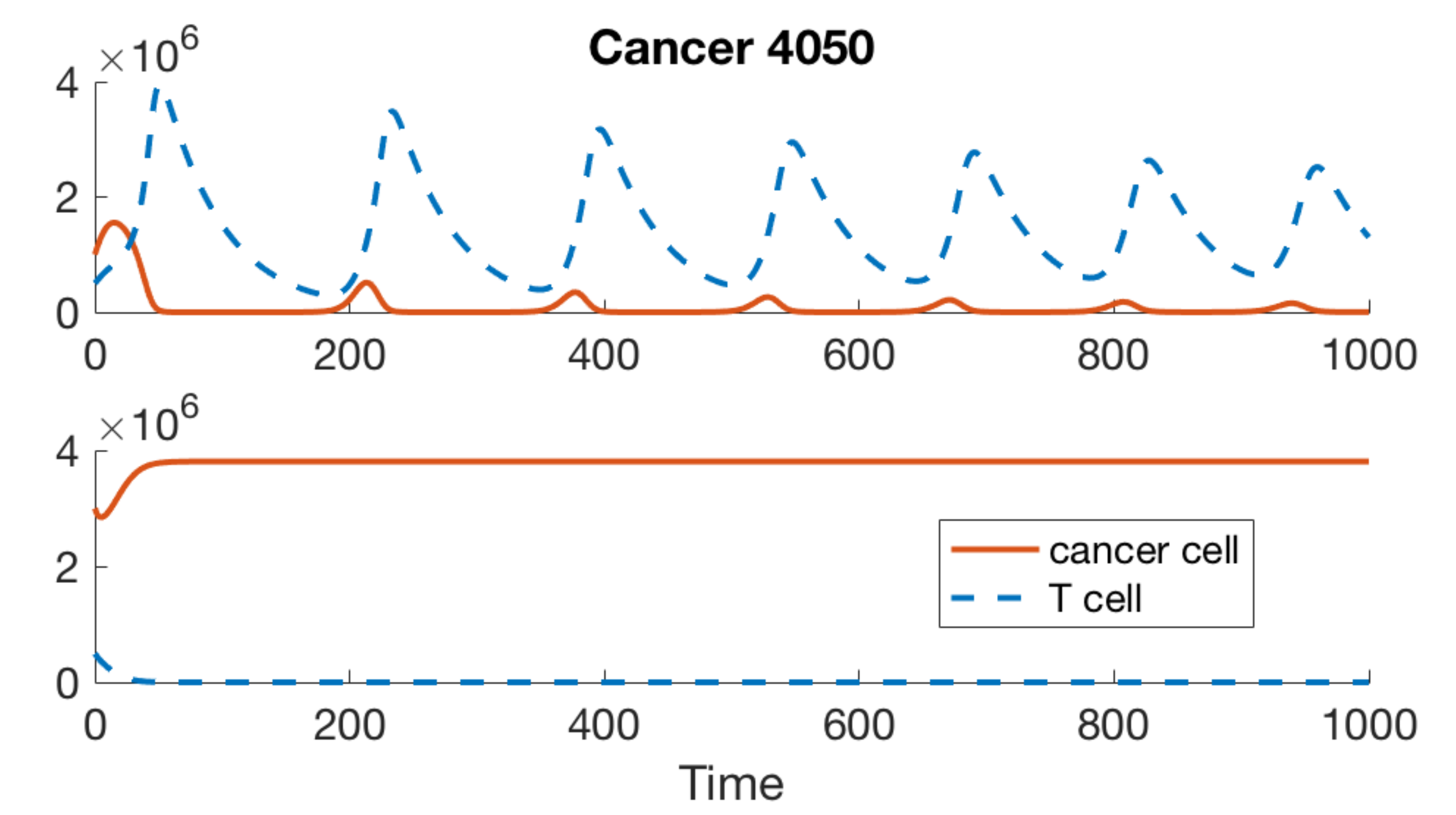}
		\includegraphics[width=4.2cm]{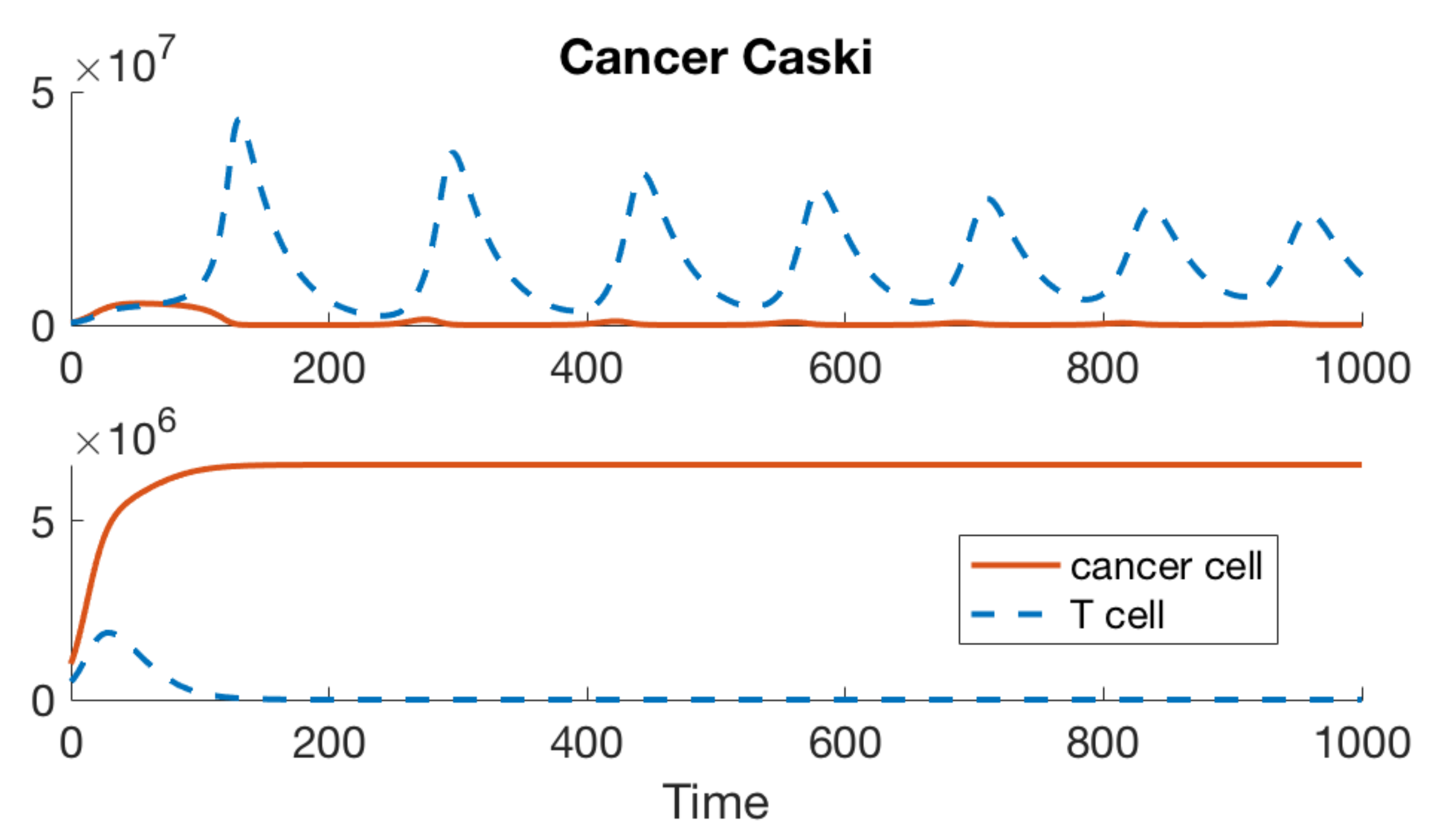}
	}
	\centerline{ 
		\includegraphics[width=3cm]{cancer_dosage_legend.pdf}
	}		
	\caption{The effect of TCR T cell therapy for different levels of initial cancer size. Treatment with $5\times 10^5$ T cells for the 4050 cell line (left) with size $1 \times 10^6$ (top) is successful, but $3 \times 10^6$ (bottom) is not. 
	For the CaSki cell line (right) and the same T cell dose, cancer of size $5 \times 10^5$ (top) can be successfully treated, but not a tumor of size $1 \times 10^6$ (bottom). 
	} 
	\label{fig:CnT_other} 
\end{figure}

The stability analysis of Section~\ref{sec:stability} suggests that if the parameters satisfy the condition in Eq.~\eqref{eq:pCond}, the system can either converge to a therapy success or failure outcome. 
We ensured that the parameters identified for the data of the 4050 cell line and the CaSki cell line in Fig.~\ref{fig:data2} fall into this category, since the data shows both trajectories depending on the initial T cell dosage. 

With the identified parameters, the phase plane of the system can provide the effective dose of T cell therapy with respect to the initial cancer size. 
Fig.~\ref{fig:phase} presents the phase plane of the 4050 and the CaSki cell lines in linear (left) and log-scale (right).  
This result provides a suggested minimum dose of T cell therapy that yields tumor reduction depending on the initial cancer size, and in fact, a therapeutic window of T cell dosages. 
In both cell lines, the smallest experimental dosage of $10^5$ falls within the range of insufficient dosage, and cancer eventually grows to its maximum capacity.  
However, the medium experimental dosage of $10^6$ is within the therapeutic window, and despite the initial increase in tumor burden in the CaSki cell line, the T cells expand and the tumor shrinks. 

To study the long-term behavior of the system, the dynamics of cancer and T cells up to 1000 days are shown in Fig.~\ref{fig:CnT_long}. 
The initial tumor size is taken at the carrying capacity, that is $T(0)=3.81\times 10^6 $ for the 4050 cell line and $6.58 \times 10^6 $ for the CaSki cell line.
The therapeutic window for this initial cancer size is $(1.3 \times 10^6, 3.8 \times 10^6)$ and $(5.2 \times 10^6, 5.0 \times 10^7)$, for the 4050 and the CaSki cell lines, respectively. 
The results shown in Fig.~\ref{fig:CnT_long} show the simulation of a TCR T cell dosage that is below, within, and above the therapeutic window. 
For the 4050 cell line, we test TCR T cells dosages of $1.2 \times 10^6 $, $3.0 \times 10^6$, and $5 \times 10^6$. 
The dosage below the window drives the tumor growth to its capacity despite its initial decline. 
On the other hand, the T cell dose within in the window effectively reduces the cancer size from $ 3.81\times 10^6 $ to $2.67 \times 10^4 $, approximately, 100 times smaller in size. 

An interesting observation is the case of a dosage that is above the therapeutic window.
In this case we observe tumor regrowth. The initial reduction of cancer is overturned and the cancer escapes the TCR T cell therapy after approximately 200 days. 
The tumor immune escape has been reported not only in an innate immune system \cite{DWherry2011}, but also in an adoptive immune system \cite{Brown2019}. 
It is presumed that an extreme does with high levels of T cells may cause premature T cell exhaustion and loss of anti-tumor activity. 

For the CaSki cell line, similar results are shown in the right column of Fig. \ref{fig:CnT_long}. A TCR T cell dosage within the range of $( 5.2 \times 10^6, 5.0 \times 10^7)$ results with a tumor reduction of approximately 65 times from $ 6.58 \times 10^6 $ to $ 1.04 \times 10^5 $. However, for other dosages, therapy fails. 
Once again we verify the effective dosage characterized in Fig.~\ref{fig:phase} by considering different initial cancer sizes. 
The results shown in Fig.~\ref{fig:CnT_other} are obtained using the T cell dosage of $5\times 10^5 $ for both cell lines, where the initial cancer size is taken as $1 \times 10^6$ and $3 \times 10^6$ for 4050, and $5 \times 10^5$ and $1 \times 10^6$ for CaSki. 
While the dosage of $5\times 10^5 $ was sufficient to reduce smaller cancers, the larger cancers cannot be reduced by this dosage. 

The results of this section stress the significance of the dosage of T cells in driving treatment success, especially given the toxicity of high-dosages. Moreover, our model can be used to identify the effective therapeutic window of T cell dosages in different cancer cell lines as a function of the initial tumor size. 
This result can potentially guide future therapy design.

\subsection{Studying the combination of T cell and IL-2 treatments, and the effect of IL-2 scheduling}
\label{sec:result3}

In addition to TCR T cell therapy, IL-2 treatment can stimulate the anti-tumor effect of TCR T cells. 
The experimental data from \cite{Jin2018} provides the IL-2 treatment for three consecutive days with dosage 198,000 IU. Jin \emph{et al.} demonstrate that the combination of TCR T cell and IL-2 treatment is especially valuable when the T cells are given at low dosages.
For instance, the IL-2 treatment did not show any apparent effect when the T cell therapy is given in high dosages of $10^7$ cells. However, it improved the T cell treatment in the 4050 cell line treated with $10^5$ cells and in the CaSki cell line treated with $10^6$ cells. 
In this section, we calibrate the model (\ref{eq:ode1}--(\ref{eq:ode3}) to the experimental data with IL-2 treatment administered at three consecutive days and study the effect of altering the treatment schedules, while keeping the total dosage administered throughout the treatment as 594,000 IU. 

\begin{figure}
	
	\centerline{ 
		\includegraphics[width=9cm]{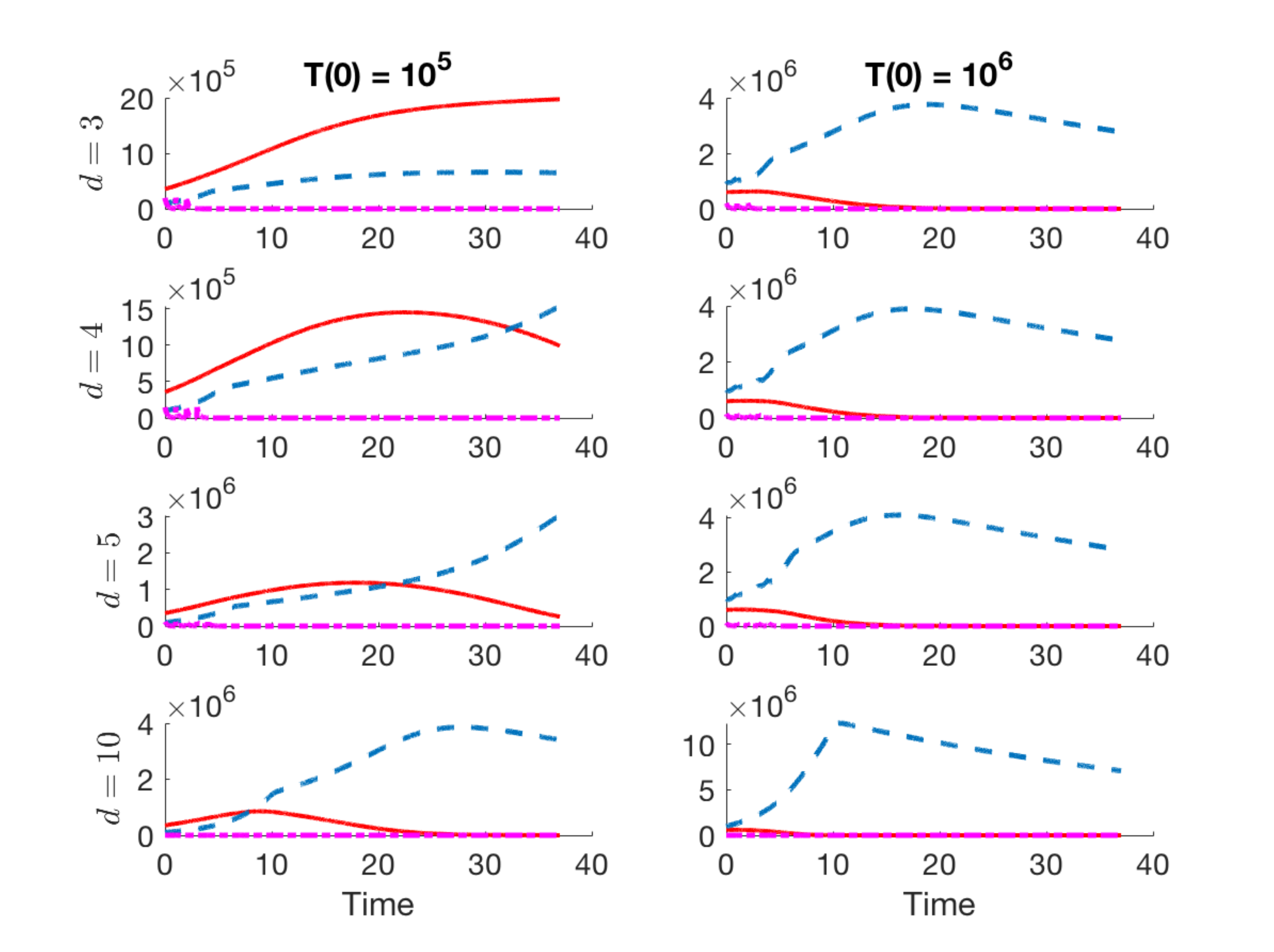}
	}
	\centerline{ 
		\includegraphics[width=5cm]{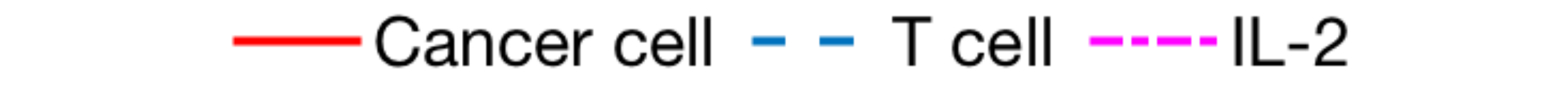}
	}	
	\caption{Combination of TCR T cell and IL-2 treatments on the 4050 cell line for different IL-2 treatment schedules. 
	The IL-2 is administered for $d=3,\,4,\,5,\,10$ days with a total dosage of $594,000 / d$ IU. The 10 days schedule shows the biggest improvement on T cell therapy. 
	} 
	\label{fig:4050IL2} 
\end{figure}

\begin{figure}
	
	\centerline{ 
		\includegraphics[width=9cm]{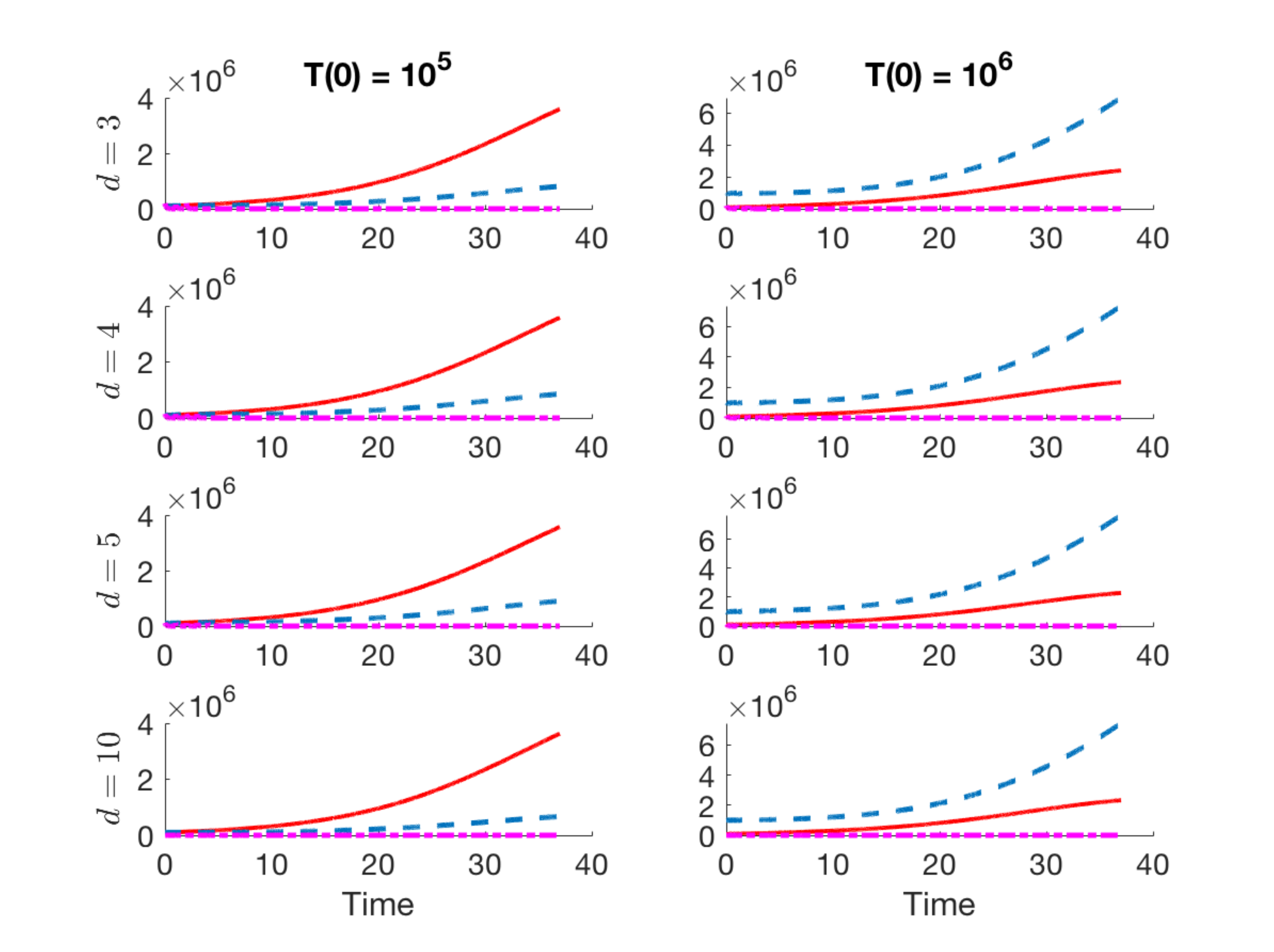}
	}
	\centerline{ 
		\includegraphics[width=5cm]{cancer_IL2_dosage_legend.pdf}
	}		
	\caption{Combination of T cell and IL-2 treatment on the CaSki cell line for different IL-2 treatment schedules. 
	The IL-2 is administered for $d=3,\,4,\,5,\,10$ days with a total dosage of $594,000 / d$ IU. In contrast to 4050 cell line, the combination therapy outcome does not depend on the IL-2 treatment schedule. 
	} 
	\label{fig:CaskiIL2} 
\end{figure}

Figures~\ref{fig:4050IL2} and~\ref{fig:CaskiIL2} present the results of alternating dosage for the 4050 and CaSki cell lines, respectively. The treatment is given for $d=3$, 4, 5, and 10 consecutive days with a total dosage of $594,000 / d$ IU. In the case of the 4050 cell line, distributing the IL-2 treatment over multiple days improves the T cell treatment of dosage $10^5$. Figure~\ref{fig:4050IL2} shows that the final tumor size is smallest when IL-2 is given for 10 days with a total dosage of $594,000$ IU. For the T cell dosage of $10^6$, the cancer shrinks in all treatment schedules. However, the T cells expand to larger magnitudes when IL-2 given for longer periods. 
On the other hand, altering the IL-2 schedule does not affect the T cell treatment outcome in the CaSki cell line as shown in Figure~\ref{fig:CaskiIL2}. The tumor size does not change despite the different IL-2 treatment schedules. The experiments in \cite{Jin2018} show that both CD8 and CD4 TCR T cells are effective for the 4050 cell line, while only CD8 TCR T cells are cytotoxic in the CaSki cell line. Although we do not model CD4 and CD8 T cells separately, our results are consistent with the experiments that show that the 4050 cell line is more affected by the T cell therapy and by the combination therapy.

\section{Conclusion}
\label{sec:conclusion}
In this paper, we study the combination of adoptive immune cell transfer therapy using E7 targeted TCR T cell and IL-2 treatment. By a sequential calibration of the model using the MCMC algorithm, we obtain the parameter values of two cancer cell lines, 4050 and CaSki, that agrees with the experimental data of \cite{Jin2018}.

We derive a condition for therapy success and failure, allowing us to study the impact of the T cell activation rate. This provides tools for calculating the minimum level of TCR T cell activation rate that is necessary for the treatment to be successful. 
When the T cell activation rate is within the range of potential therapy success, we obtain a therapeutic window for the T cell dose as a function of the tumor size. The results are verified numerically for both cell lines. 
This emphasizes that the tumor size should be taken into account when deciding the dosage, in addition to the general practice that is based on the weight of the patient. Moreover, the model illustrates the scenario of toxicity with a high-dosage of T cell therapy, and treatment failure after transient regression that has been observed in the adoptive cell therapy community. 
Finally, the combination of TCR T cell and IL-2 treatment is studied, where we demonstrate that modifying the treatment schedule of IL-2 can potentially improve treating the 4050 cell line, but not the CaSki cell line. 

Our future work includes modeling distinct types of TCR T cells and immune cells, such as effector, helper, regulatory, and memory T cells. In addition, we propose to model the interaction between the immune cells and different cytokines including IL-2 that can help us understand the complex dynamics of the immune system and make robust predictions regarding the expected outcomes of immunotherapy \citep{Paucek2019}. In particular, the cytokine release syndrome and neurologic toxicities are major side effects of adoptive T cell therapy 
for which mathematical models can provide insights given 
the lack of informative animal models. 
To improve T cell therapy, we propose to model receptor density and to understand T cell exhaustion and its effect on off-target cells \citep{Long2015}. 


\appendix 

\section{Steady states} 
\label{sec:Apendix2}

\textbf{ Theorem 1.} {\em T cell therapy fails regardless of the dose if $p < (\sqrt{d} + \sqrt{mg})^2$. The therapy succeeds if $ (\frac{m}{b} +d)(gb+1)< p$. 
If the T cell proliferation is in the range of  $ (\sqrt{d} + \sqrt{mg})^2 < p < (\frac{m}{b} +d)(gb+1) $, the treatment success depends on the initial cancer size and the T cell dosage. 
}

\begin{pf}
	\begin{enumerate}
		\item $T = 0$ and $C = 0$. $ (T,C) = (0,0) $ is a saddle, since the linearized Jacobian reduces to 
			$ \begin{pmatrix}
		-d  & 0\\
		0  & a
		\end{pmatrix}
		$. 		
This is a trivial equilibrium state, and at this state, there is no tumor and no T cells. 
		
		\item $T = 0$ and $a(1-bC)-nT = 0$. Plugging in we have $C = b^{-1} $ and $ T = 0$. $ (T,C) = (0,b^{-1}) $  is an equilibrium state where tumor cells reach maximum capacity, while T cells are absent. The linearized Jacobian reduces to 
			$$ \begin{pmatrix}
		\displaystyle{-d + \frac{p}{gb+1} - m/b} & 0\\
		-n/b  & -a
		\end{pmatrix}$$
		This point becomes stable when $-d + \frac{p}{gb+1} - m/b < 0$ which holds if $ p < \left(\frac{m}{b} +d\right)(gb+1)  $, otherwise becomes unstable. 
			
		\item $a(1-bC)-nT = 0$ and $- d + p\frac{C}{g+C} - mC = 0$. Rearranging we have two equilibrium points $$C = \frac{(p-d-mg) \pm \sqrt {(p-d-mg)^2-4mgd}}{2m}$$ and  $$ T = \frac{a(1-bC)}{n}$$
		so to have real stationary points, we must have $(p-d-mg)^2-4mgd \geq 0$.
		
		\begin{enumerate}
			\item $p-d-mg \geq 0$, then we have $ p \geq d+mg+2\sqrt{mgd} = (\sqrt{d} + \sqrt{mg})^2  $ and two positive $C$, denoted $ 0 < C_1 < C_2$. 
			\item $p-d-mg < 0$, then we have $  p \leq d+mg-2\sqrt{mgd} = (\sqrt{d} - \sqrt{mg})^2 $ and two negative $C$.  
		\end{enumerate}
		
		 Note that if we want $ C $ to be non-negative, then condition (a) must hold, i.e, $ p \geq (\sqrt{d} + \sqrt{mg})^2$, which is also consistent with the biological fact that compared to the apoptosis and the death rate due to competition, the proliferation or activation rate of T cells must be at least in the same level. Otherwise tumor cells will reach their maximum capacity $b^{-1}$. 
		
		To check for stability, let $ g(C)  = - d + p\frac{C}{g+C} - mC $ and equilibrium state $(T,C) = ( T_i, C_i )$ with $ i = 1, 2 $. Then  we have $ T_1 > T_2 > 0 $, $g'(C_1) > 0$, $g'(C_2) < 0$, and the linearized Jacobian reduces to
		$$L = \begin{pmatrix}
		0 & g'(C_i)T_i\\
		-nC_i  & -abC_i
		\end{pmatrix},$$ 
		$$P_L(\lambda) = \lambda^2 + (abC_i)\lambda + nC_ig'(C_i)T_i$$ 
		
		The eigenvalues are $$\lambda _{1,2} = \frac{-(abC_i) \pm \sqrt{(abC_i)^2 - 4nC_ig'(C_i)T_i} }{2}.$$  
		
		Since $g'(C_1) > 0$, $(T_1, C_1)$ is a stable nodal sink if $ (abC_i)^2 - 4nC_ig'(C_i)T_i > 0 $, a stable twist sink if $ (abC_i)^2 - 4nC_ig'(C_i)T_i = 0 $, and a stable spiral sink if $ (abC_i)^2 - 4nC_ig'(C_i)T_i < 0 $. Since $g'(C_2) < 0$, $(T_2, C_2)$ is a saddle.
    \end{enumerate}
$ $ 
  
In short, the range in terms of T cell proliferation can be ordered as 
$$ d+mg \leq d+mg+2\sqrt{mgd} \leq \left(\frac{m}{b} +d\right)(gb+1),$$
\indent which classifies the stability of the equilibrium points. 
\qed
\end{pf}

\bibliographystyle{cas-model2-names}

\bibliography{references}

\begin{thebibliography}{45}
\expandafter\ifx\csname natexlab\endcsname\relax\def\natexlab#1{#1}\fi
\providecommand{\url}[1]{\texttt{#1}}
\providecommand{\href}[2]{#2}
\providecommand{\path}[1]{#1}
\providecommand{\DOIprefix}{doi:}
\providecommand{\ArXivprefix}{arXiv:}
\providecommand{\URLprefix}{URL: }
\providecommand{\Pubmedprefix}{pmid:}
\providecommand{\doi}[1]{\href{http://dx.doi.org/#1}{\path{#1}}}
\providecommand{\Pubmed}[1]{\href{pmid:#1}{\path{#1}}}
\providecommand{\bibinfo}[2]{#2}
\ifx\xfnm\relax \def\xfnm[#1]{\unskip,\space#1}\fi
\bibitem[{Barber et~al.(2006))Barber, Wherry, Masopust, Zhu, Allison, Sharpe,
  Freeman and Ahmed}]{Barber2006}
\bibinfo{author}{Barber, D.L.}, \bibinfo{author}{Wherry, E.J.},
  \bibinfo{author}{Masopust, D.}, \bibinfo{author}{Zhu, B.},
  \bibinfo{author}{Allison, J.P.}, \bibinfo{author}{Sharpe, A.H.},
  \bibinfo{author}{Freeman, G.J.}, \bibinfo{author}{Ahmed, R.},
  \bibinfo{year}{2006)}.
\newblock \bibinfo{title}{Restoring function in exhausted cd8 t cells during
  chronic viral infection}.
\newblock \bibinfo{journal}{Nature} \bibinfo{volume}{439},
  \bibinfo{pages}{682–687}.
\bibitem[{Benmebarek et~al.(2019)Benmebarek, Karches, Cadilha, Lesch, Endres
  and Kobold}]{Benmebarek2019}
\bibinfo{author}{Benmebarek, M.R.}, \bibinfo{author}{Karches, C.H.},
  \bibinfo{author}{Cadilha, B.L.}, \bibinfo{author}{Lesch, S.},
  \bibinfo{author}{Endres, S.}, \bibinfo{author}{Kobold, S.},
  \bibinfo{year}{2019}.
\newblock \bibinfo{title}{{Killing mechanisms of chimeric antigen receptor
  (CAR) T cells}}.
\newblock \bibinfo{journal}{International Journal of Molecular Sciences}
  \bibinfo{volume}{20}, \bibinfo{pages}{1--21}.
\bibitem[{Brown and Mackall(2019)}]{Brown2019}
\bibinfo{author}{Brown, C.E.}, \bibinfo{author}{Mackall, C.L.},
  \bibinfo{year}{2019}.
\newblock \bibinfo{title}{Car t cell therapy: inroads to response and
  resistance}.
\newblock \bibinfo{journal}{Nature Reviews Immunology} \bibinfo{volume}{19},
  \bibinfo{pages}{73--74}.
\bibitem[{Busse et~al.(2010)Busse, {De La Rosa}, Hobiger, Thurley, Flossdorf,
  Scheffold and H{\"{o}}fer}]{Busse2010}
\bibinfo{author}{Busse, D.}, \bibinfo{author}{{De La Rosa}, M.},
  \bibinfo{author}{Hobiger, K.}, \bibinfo{author}{Thurley, K.},
  \bibinfo{author}{Flossdorf, M.}, \bibinfo{author}{Scheffold, A.},
  \bibinfo{author}{H{\"{o}}fer, T.}, \bibinfo{year}{2010}.
\newblock \bibinfo{title}{{Competing feedback loops shape IL-2 signaling
  between helper and regulatory T lymphocytes in cellular microenvironments}}.
\newblock \bibinfo{journal}{PNAS} \bibinfo{volume}{107},
  \bibinfo{pages}{3058--3063}.
\bibitem[{{De Pillis} and Radunskaya(2003)}]{DePillis2003}
\bibinfo{author}{{De Pillis}, L.G.}, \bibinfo{author}{Radunskaya, A.},
  \bibinfo{year}{2003}.
\newblock \bibinfo{title}{{The dynamics of an optimally controlled tumor model:
  A case study}}.
\newblock \bibinfo{journal}{Mathematical and Computer Modelling}
  \bibinfo{volume}{37}, \bibinfo{pages}{1221--1244}.
\bibitem[{D'Onofrio(2008)}]{DOnofrio2008}
\bibinfo{author}{D'Onofrio, A.}, \bibinfo{year}{2008}.
\newblock \bibinfo{title}{{Metamodeling tumor-immune system interaction, tumor
  evasion and immunotherapy}}.
\newblock \bibinfo{journal}{Mathematical and Computer Modelling}
  \bibinfo{volume}{47}, \bibinfo{pages}{614--637}.
\bibitem[{Dudley et~al.(2003)Dudley, Wunderlich, Shelton, Even and
  Rosenberg}]{Dudley2003}
\bibinfo{author}{Dudley, M.E.}, \bibinfo{author}{Wunderlich, J.R.},
  \bibinfo{author}{Shelton, T.E.}, \bibinfo{author}{Even, J.},
  \bibinfo{author}{Rosenberg, S.A.}, \bibinfo{year}{2003}.
\newblock \bibinfo{title}{Generation of tumor-infiltrating lymphocyte cultures
  for use in adoptive transfer therapy for melanoma patients}.
\newblock \bibinfo{journal}{J. Immunother.} \bibinfo{volume}{26},
  \bibinfo{pages}{332--342}.
\bibitem[{F.~Castiglione(2007)}]{Piccoli2007}
\bibinfo{author}{F.~Castiglione, B.P.}, \bibinfo{year}{2007}.
\newblock \bibinfo{title}{Cancer immunotherapy, mathematical modeling and
  optimal control}.
\newblock \bibinfo{journal}{J. Theor. Biol.} \bibinfo{volume}{247},
  \bibinfo{pages}{723--732}.
\bibitem[{Govers et~al.(2010)Govers, Sebestyen, Coccoris, Willemsen and
  Debets}]{Govers2010}
\bibinfo{author}{Govers, C.}, \bibinfo{author}{Sebestyen, Z.},
  \bibinfo{author}{Coccoris, M.}, \bibinfo{author}{Willemsen, R.A.},
  \bibinfo{author}{Debets, R.}, \bibinfo{year}{2010}.
\newblock \bibinfo{title}{T cell receptor gene therapy: strategies for
  optimizing transgenic {TCR} pairing}.
\newblock \bibinfo{journal}{Trends in Molecular Medicine} \bibinfo{volume}{16},
  \bibinfo{pages}{77--87}.
\bibitem[{Haario et~al.(2006)Haario, Laine, Mira and Saksman}]{Haario2006}
\bibinfo{author}{Haario, H.}, \bibinfo{author}{Laine, M.},
  \bibinfo{author}{Mira, A.}, \bibinfo{author}{Saksman, E.},
  \bibinfo{year}{2006}.
\newblock \bibinfo{title}{{DRAM}: Efficient adaptive {MCMC}}.
\newblock \bibinfo{journal}{Statistics and Computing} \bibinfo{volume}{16},
  \bibinfo{pages}{339--354}.
\bibitem[{Hardiansyah and Ng(2019)}]{Hardiansyah2019}
\bibinfo{author}{Hardiansyah, D.}, \bibinfo{author}{Ng, C.M.},
  \bibinfo{year}{2019}.
\newblock \bibinfo{title}{Pharmacology model of chimeric antigen receptor
  {T}-cell therapy}.
\newblock \bibinfo{journal}{Clin. Transl. Sci.} \bibinfo{volume}{12},
  \bibinfo{pages}{343--349}.
\bibitem[{Hinrichs(2016)}]{Hinrichs2016}
\bibinfo{author}{Hinrichs, C.S.}, \bibinfo{year}{2016}.
\newblock \bibinfo{title}{Molecular pathways: Breaking the epithelial cancer
  barrier for chimeric antigen receptor and {T}-cell receptor gene therapy}.
\newblock \bibinfo{journal}{Clin Cancer Res.} \bibinfo{volume}{22},
  \bibinfo{pages}{1559--1564}.
\bibitem[{Hopkins et~al.(2018)Hopkins, Tucker, Pan, Fang and
  Huang}]{Hopkins2018}
\bibinfo{author}{Hopkins, B.}, \bibinfo{author}{Tucker, M.},
  \bibinfo{author}{Pan, Y.}, \bibinfo{author}{Fang, N.},
  \bibinfo{author}{Huang, Z.}, \bibinfo{year}{2018}.
\newblock \bibinfo{title}{A model-based investigation of cytokine storm for
  {T}-cell therapy}.
\newblock \bibinfo{journal}{IFAC-PapersOnLine} \bibinfo{volume}{51},
  \bibinfo{pages}{76--79}.
\bibitem[{Jin et~al.(2018)Jin, Campbell, Draper, Stevanovi{\'{c}}, Weissbrich,
  Yu, Restifo, Rosenberg, Trimble and Hinrichs}]{Jin2018}
\bibinfo{author}{Jin, B.Y.}, \bibinfo{author}{Campbell, T.E.},
  \bibinfo{author}{Draper, L.M.}, \bibinfo{author}{Stevanovi{\'{c}}, S.},
  \bibinfo{author}{Weissbrich, B.}, \bibinfo{author}{Yu, Z.},
  \bibinfo{author}{Restifo, N.P.}, \bibinfo{author}{Rosenberg, S.A.},
  \bibinfo{author}{Trimble, C.L.}, \bibinfo{author}{Hinrichs, C.S.},
  \bibinfo{year}{2018}.
\newblock \bibinfo{title}{{Engineered T cells targeting E7 mediate regression
  of human papillomavirus cancers in a murine model}}.
\newblock \bibinfo{journal}{JCI insight} \bibinfo{volume}{3},
  \bibinfo{pages}{1--12}.
\bibitem[{June et~al.(2018)June, O'Connor, Kawalekar, Ghassemi and
  Milone}]{June2018}
\bibinfo{author}{June, C.H.}, \bibinfo{author}{O'Connor, R.S.},
  \bibinfo{author}{Kawalekar, O.U.}, \bibinfo{author}{Ghassemi, S.},
  \bibinfo{author}{Milone, M.C.}, \bibinfo{year}{2018}.
\newblock \bibinfo{title}{{CAR T cell immunotherapy for human cancer}}.
\newblock \bibinfo{journal}{Science} \bibinfo{volume}{359},
  \bibinfo{pages}{1361--1365}.
\bibitem[{Kim et~al.(2008)Kim, Lee and Levy}]{PKim2008}
\bibinfo{author}{Kim, P.}, \bibinfo{author}{Lee, P.}, \bibinfo{author}{Levy,
  D.}, \bibinfo{year}{2008}.
\newblock \bibinfo{title}{Dynamics and potential impact of the immune response
  to chronic myelogenous leukemia}.
\newblock \bibinfo{journal}{PLOS Comput. Biol.} \bibinfo{volume}{4},
  \bibinfo{pages}{1--17}.
\bibitem[{Kirschner and Panetta(1998)}]{Kirschner1998}
\bibinfo{author}{Kirschner, D.}, \bibinfo{author}{Panetta, J.C.},
  \bibinfo{year}{1998}.
\newblock \bibinfo{title}{Modeling immunotherapy of the tumor-immune
  interaction}.
\newblock \bibinfo{journal}{J. Math. Biol.} \bibinfo{volume}{37},
  \bibinfo{pages}{235--252}.
\bibitem[{Konstorum et~al.(2017)Konstorum, Vella, Adler and
  Laubenbacher}]{Konstorum2017}
\bibinfo{author}{Konstorum, A.}, \bibinfo{author}{Vella, A.T.},
  \bibinfo{author}{Adler, A.J.}, \bibinfo{author}{Laubenbacher, R.C.},
  \bibinfo{year}{2017}.
\newblock \bibinfo{title}{{Addressing current challenges in cancer
  immunotherapy with mathematical and computational modelling}}.
\newblock \bibinfo{journal}{Journal of the Royal Society Interface}
  \bibinfo{volume}{14}, \bibinfo{pages}{1--10}.
\bibitem[{Kurachi(2019)}]{Kurachi2019}
\bibinfo{author}{Kurachi, M.}, \bibinfo{year}{2019}.
\newblock \bibinfo{title}{{CD8+ T} cell exhaustion}.
\newblock \bibinfo{journal}{Seminars in Immunopathology} \bibinfo{volume}{41},
  \bibinfo{pages}{327–337}.
\bibitem[{Kuznetsov et~al.(1994)Kuznetsov, Makalkin, Taylor and
  Perelson}]{Kuznetsov1994}
\bibinfo{author}{Kuznetsov, V.A.}, \bibinfo{author}{Makalkin, I.A.},
  \bibinfo{author}{Taylor, M.A.}, \bibinfo{author}{Perelson, A.S.},
  \bibinfo{year}{1994}.
\newblock \bibinfo{title}{Nonlinear dynamics of immunogen1c tumors: Parameter
  estimation and global bifurcation analysis}.
\newblock \bibinfo{journal}{Bull. Math. Biol.} \bibinfo{volume}{56},
  \bibinfo{pages}{295--321}.
\bibitem[{Long et~al.(2015)Long, Haso, Shern, Wanhainen, Murgai, Ingaramo,
  Smith, Walker, Kohler, Venkateshwara, Kaplan, Patterson, Fry, Orentas and
  Mackall}]{Long2015}
\bibinfo{author}{Long, A.H.}, \bibinfo{author}{Haso, W.M.},
  \bibinfo{author}{Shern, J.F.}, \bibinfo{author}{Wanhainen, K.M.},
  \bibinfo{author}{Murgai, M.}, \bibinfo{author}{Ingaramo, M.},
  \bibinfo{author}{Smith, J.P.}, \bibinfo{author}{Walker, A.J.},
  \bibinfo{author}{Kohler, M.E.}, \bibinfo{author}{Venkateshwara, V.R.},
  \bibinfo{author}{Kaplan, R.N.}, \bibinfo{author}{Patterson, G.H.},
  \bibinfo{author}{Fry, T.J.}, \bibinfo{author}{Orentas, R.J.},
  \bibinfo{author}{Mackall, C.L.}, \bibinfo{year}{2015}.
\newblock \bibinfo{title}{{4-1BB} costimulation ameliorates {T} cell exhaustion
  induced by tonic signaling of chimeric antigen receptors}.
\newblock \bibinfo{journal}{Nature Medicine} \bibinfo{volume}{21},
  \bibinfo{pages}{581--590}.
\bibitem[{Lynn et~al.(2019)Lynn, Weber, Gennert, Sotillo, Xu, Good, Anbunathan,
  Jones, Tieu, Granja, DeBourcy, Majzner, Satpathy, Quake, Chang and
  Mackall}]{Lynn2019}
\bibinfo{author}{Lynn, R.C.}, \bibinfo{author}{Weber, E.W.},
  \bibinfo{author}{Gennert, D.}, \bibinfo{author}{Sotillo, E.},
  \bibinfo{author}{Xu, P.}, \bibinfo{author}{Good, Z.},
  \bibinfo{author}{Anbunathan, H.}, \bibinfo{author}{Jones, R.},
  \bibinfo{author}{Tieu, V.}, \bibinfo{author}{Granja, J.},
  \bibinfo{author}{DeBourcy, C.}, \bibinfo{author}{Majzner, R.},
  \bibinfo{author}{Satpathy, A.T.}, \bibinfo{author}{Quake, S.R.},
  \bibinfo{author}{Chang, H.}, \bibinfo{author}{Mackall, C.L.},
  \bibinfo{year}{2019}.
\newblock \bibinfo{title}{{c-Jun Overexpressing CAR-T Cells are
  Exhaustion-Resistant and Mediate Enhanced Antitumor Activity}}.
\newblock \bibinfo{journal}{bioRxiv} , \bibinfo{pages}{1--33}.
\bibitem[{McLane et~al.(2019)McLane, Abdel-Hakeem and Wherry}]{McLane2019}
\bibinfo{author}{McLane, L.M.}, \bibinfo{author}{Abdel-Hakeem, M.S.},
  \bibinfo{author}{Wherry, E.J.}, \bibinfo{year}{2019}.
\newblock \bibinfo{title}{{CD8 T} cell exhaustion during chronic viral
  infection and cancer}.
\newblock \bibinfo{journal}{Annual Review of Immunology} \bibinfo{volume}{37},
  \bibinfo{pages}{457--495}.
\bibitem[{Moore and Li(2004)}]{Moore2004}
\bibinfo{author}{Moore, H.}, \bibinfo{author}{Li, N.K.}, \bibinfo{year}{2004}.
\newblock \bibinfo{title}{{A mathematical model for chronic myelogenous
  leukemia (CML) and T cell interaction}}.
\newblock \bibinfo{journal}{Journal of Theoretical Biology}
  \bibinfo{volume}{227}, \bibinfo{pages}{513--523}.
\bibitem[{Mostolizadeh et~al.(2018)Mostolizadeh, Afsharnezhad and
  Marciniak-Czochra}]{Mostolizadeh2018}
\bibinfo{author}{Mostolizadeh, R.}, \bibinfo{author}{Afsharnezhad, Z.},
  \bibinfo{author}{Marciniak-Czochra, A.}, \bibinfo{year}{2018}.
\newblock \bibinfo{title}{Mathematical model of chimeric anti-gene receptor
  {(CAR) T} cell therapy with presence of cytokine}.
\newblock \bibinfo{journal}{Numerical Algebra, Control \& Optimization}
  \bibinfo{volume}{8}, \bibinfo{pages}{63--80}.
\bibitem[{Nikolopoulou et~al.(2018)Nikolopoulou, Johnson, Harris, Nagy, Stites
  and Kuang}]{Nikolopoulou2018}
\bibinfo{author}{Nikolopoulou, E.}, \bibinfo{author}{Johnson, L.R.},
  \bibinfo{author}{Harris, D.}, \bibinfo{author}{Nagy, J.D.},
  \bibinfo{author}{Stites, E.C.}, \bibinfo{author}{Kuang, Y.},
  \bibinfo{year}{2018}.
\newblock \bibinfo{title}{{Tumour-immune dynamics with an immune checkpoint
  inhibitor}}.
\newblock \bibinfo{journal}{Letters in Biomathematics} \bibinfo{volume}{5},
  \bibinfo{pages}{S137--S159}.
\bibitem[{Paucek et~al.(2019)Paucek, Baltimore and Li}]{Paucek2019}
\bibinfo{author}{Paucek, D.}, \bibinfo{author}{Baltimore, D.},
  \bibinfo{author}{Li, G.}, \bibinfo{year}{2019}.
\newblock \bibinfo{title}{The cellular immunotherapy revolution: Arming the
  immune system for precision therapy}.
\newblock \bibinfo{journal}{Trends in Immunology} \bibinfo{volume}{40},
  \bibinfo{pages}{292--309}.
\bibitem[{Peskov et~al.(2019)Peskov, Azarov, Chu, Voronova, Kosinsky and
  Helmlinger}]{Peskov2019}
\bibinfo{author}{Peskov, K.}, \bibinfo{author}{Azarov, I.},
  \bibinfo{author}{Chu, L.}, \bibinfo{author}{Voronova, V.},
  \bibinfo{author}{Kosinsky, Y.}, \bibinfo{author}{Helmlinger, G.},
  \bibinfo{year}{2019}.
\newblock \bibinfo{title}{{Quantitative mechanistic modeling in support of
  pharmacological therapeutics development in immuno-oncology}}.
\newblock \bibinfo{journal}{Frontiers in Immunology} \bibinfo{volume}{10},
  \bibinfo{pages}{1--11}.
\bibitem[{de~Pillis(2007)}]{DePillis2007}
\bibinfo{author}{de~Pillis, L.G.}, \bibinfo{year}{2007}.
\newblock \bibinfo{title}{Chemotherapy for tumors: An analysis of the dynamics
  and a study of quadratic and linear optimal controls}.
\newblock \bibinfo{journal}{Mathematical Biosciences} \bibinfo{volume}{209},
  \bibinfo{pages}{292--315}.
\bibitem[{de~Pillis et~al.(2009)de~Pillis, Fister, Gu, Collins, Daub, Gross,
  Moore and Preskill}]{DePillis2009}
\bibinfo{author}{de~Pillis, L.G.}, \bibinfo{author}{Fister, K.R.},
  \bibinfo{author}{Gu, W.}, \bibinfo{author}{Collins, C.},
  \bibinfo{author}{Daub, M.}, \bibinfo{author}{Gross, D.},
  \bibinfo{author}{Moore, J.}, \bibinfo{author}{Preskill, B.},
  \bibinfo{year}{2009}.
\newblock \bibinfo{title}{{Mathematical model creation for cancer
  chemo-immunotherapy}}.
\newblock \bibinfo{journal}{Computational and Mathematical Methods in Medicine}
  \bibinfo{volume}{10}, \bibinfo{pages}{165--184}.
\bibitem[{de~Pillis et~al.(2006)de~Pillis, Gu and Radunskaya}]{DePillis2006}
\bibinfo{author}{de~Pillis, L.G.}, \bibinfo{author}{Gu, W.},
  \bibinfo{author}{Radunskaya, A.E.}, \bibinfo{year}{2006}.
\newblock \bibinfo{title}{Mixed immunotherapy and chemotherapy of tumors:
  Modeling, applications and biological interpretations}.
\newblock \bibinfo{journal}{J. Theor. Biol.} \bibinfo{volume}{238},
  \bibinfo{pages}{841--862}.
\bibitem[{Piotrowska(2016)}]{Piotrowska2016}
\bibinfo{author}{Piotrowska, M.J.}, \bibinfo{year}{2016}.
\newblock \bibinfo{title}{An immune system-tumour interactions model with
  discrete time delay: Model analysis and validation}.
\newblock \bibinfo{journal}{Communications in Nonlinear Science and Numerical
  Simulation} \bibinfo{volume}{34}, \bibinfo{pages}{185--198}.
\bibitem[{Radunskaya et~al.(2018)Radunskaya, Kim and {Woods
  II}}]{Radunskaya2018}
\bibinfo{author}{Radunskaya, A.}, \bibinfo{author}{Kim, R.},
  \bibinfo{author}{{Woods II}, T.}, \bibinfo{year}{2018}.
\newblock \bibinfo{title}{{Mathematical Modeling of Tumor Immune Interactions:
  a Closer Look at the Role of a PD-L1 Inhibitor in Cancer Immunotherapy}}.
\newblock \bibinfo{journal}{SPORA: A Journal of Biomathematics}
  \bibinfo{volume}{4}, \bibinfo{pages}{25--41}.
\bibitem[{Rihan et~al.(2014)Rihan, Abdel~Rahman, Lakshmanan and
  Alkhajeh}]{Rihan2014}
\bibinfo{author}{Rihan, F.A.}, \bibinfo{author}{Abdel~Rahman, D.H.},
  \bibinfo{author}{Lakshmanan, S.}, \bibinfo{author}{Alkhajeh, A.S.},
  \bibinfo{year}{2014}.
\newblock \bibinfo{title}{A time delay model of tumour–immune system
  interactions: global dynamics, parameter estimation, sensitivity analysis}.
\newblock \bibinfo{journal}{Appl. Math. Comput.} \bibinfo{volume}{232},
  \bibinfo{pages}{606--623}.
\bibitem[{Rihan et~al.(2019)Rihan, Lakshmanan and Maurer}]{Rihan2019}
\bibinfo{author}{Rihan, F.A.}, \bibinfo{author}{Lakshmanan, S.},
  \bibinfo{author}{Maurer, H.}, \bibinfo{year}{2019}.
\newblock \bibinfo{title}{Optimal control of tumour-immune model with
  time-delay and immuno-chemotherapy}.
\newblock \bibinfo{journal}{Appl. Math. Comput.} \bibinfo{volume}{353},
  \bibinfo{pages}{147--165}.
\bibitem[{Rohaan et~al.(2018)Rohaan, van~den Berg, Kvistborg and
  Haanen}]{Rohaan2018}
\bibinfo{author}{Rohaan, M.W.}, \bibinfo{author}{van~den Berg, J.H.},
  \bibinfo{author}{Kvistborg, P.}, \bibinfo{author}{Haanen, J.B.A.G.},
  \bibinfo{year}{2018}.
\newblock \bibinfo{title}{Adoptive transfer of tumor-infiltrating lymphocytes
  in melanoma: a viable treatment option}.
\newblock \bibinfo{journal}{Journal for Immunotherapy of Cancer}
  \bibinfo{volume}{6}, \bibinfo{pages}{1--16}.
\bibitem[{Sahoo et~al.(2019)Sahoo, Yang, Abler, Maestrini, Adhikarla,
  Frankhouser, Cho, Machuca, Wang, Barish, Gutova, Branciamore, Brown and
  Rockne}]{Sahoo2019}
\bibinfo{author}{Sahoo, P.}, \bibinfo{author}{Yang, X.},
  \bibinfo{author}{Abler, D.}, \bibinfo{author}{Maestrini, D.},
  \bibinfo{author}{Adhikarla, V.}, \bibinfo{author}{Frankhouser, D.},
  \bibinfo{author}{Cho, H.}, \bibinfo{author}{Machuca, V.},
  \bibinfo{author}{Wang, D.}, \bibinfo{author}{Barish, M.},
  \bibinfo{author}{Gutova, M.}, \bibinfo{author}{Branciamore, S.},
  \bibinfo{author}{Brown, C.E.}, \bibinfo{author}{Rockne, R.C.},
  \bibinfo{year}{2019}.
\newblock \bibinfo{title}{{A mathematical modeling approach to explore kinetics
  of Chimeric Antigen Receptor (CAR) T-cell Response in glioma: the CARRGO
  model}}.
\newblock \bibinfo{journal}{bioRxiv}
  \href{http://arxiv.org/abs/10.1101/786020}{\tt arXiv:10.1101/786020}.
\bibitem[{Scott et~al.(2019)Scott, Dündar, Zumbo, Chandran, Klebanoff,
  Shakiba, Prerak~Trivedi, Appleby, Camara, Zamarin, Walther, Snyder, Matthew
  R.~Femia, Wen, Matthew D.~Hellmann, Liu, Altorki, Lauer, Levy, Glickman,
  Kaye, Betel, Philip and Schietinger}]{Scott2019}
\bibinfo{author}{Scott, A.C.}, \bibinfo{author}{Dündar, F.},
  \bibinfo{author}{Zumbo, P.}, \bibinfo{author}{Chandran, S.S.},
  \bibinfo{author}{Klebanoff, C.A.}, \bibinfo{author}{Shakiba, M.},
  \bibinfo{author}{Prerak~Trivedi, a.L.M.}, \bibinfo{author}{Appleby, H.},
  \bibinfo{author}{Camara, S.}, \bibinfo{author}{Zamarin, D.},
  \bibinfo{author}{Walther, T.}, \bibinfo{author}{Snyder, A.},
  \bibinfo{author}{Matthew R.~Femia, a.E.A.C.}, \bibinfo{author}{Wen, H.Y.},
  \bibinfo{author}{Matthew D.~Hellmann, a.N.A.}, \bibinfo{author}{Liu, Y.},
  \bibinfo{author}{Altorki, N.K.}, \bibinfo{author}{Lauer, P.},
  \bibinfo{author}{Levy, O.}, \bibinfo{author}{Glickman, M.S.},
  \bibinfo{author}{Kaye, J.}, \bibinfo{author}{Betel, D.},
  \bibinfo{author}{Philip, M.}, \bibinfo{author}{Schietinger, A.},
  \bibinfo{year}{2019}.
\newblock \bibinfo{title}{{TOX} is a critical regulator of tumour-specific {T}
  cell differentiation}.
\newblock \bibinfo{journal}{Nature} \bibinfo{volume}{571},
  \bibinfo{pages}{270–274}.
\bibitem[{Seo et~al.(2019)Seo, Chen, González-Avalos, Samaniego-Castruita,
  Das, Wang, López-Moyado, Georges, Zhang, Onodera, Wu, Lu, Hogan, Bhandoola
  and Rao}]{Seo2019}
\bibinfo{author}{Seo, H.}, \bibinfo{author}{Chen, J.},
  \bibinfo{author}{González-Avalos, E.}, \bibinfo{author}{Samaniego-Castruita,
  D.}, \bibinfo{author}{Das, A.}, \bibinfo{author}{Wang, Y.H.},
  \bibinfo{author}{López-Moyado, I.F.}, \bibinfo{author}{Georges, R.O.},
  \bibinfo{author}{Zhang, W.}, \bibinfo{author}{Onodera, A.},
  \bibinfo{author}{Wu, C.J.}, \bibinfo{author}{Lu, L.F.},
  \bibinfo{author}{Hogan, P.G.}, \bibinfo{author}{Bhandoola, A.},
  \bibinfo{author}{Rao, A.}, \bibinfo{year}{2019}.
\newblock \bibinfo{title}{{TOX and TOX2 transcription factors cooperate with
  NR4A transcription factors to impose CD8+ T cell exhaustion}}.
\newblock \bibinfo{journal}{PNAS} \bibinfo{volume}{116},
  \bibinfo{pages}{12410--12415}.
\bibitem[{Sharma and Allison(2015)}]{Sharma2015}
\bibinfo{author}{Sharma, P.}, \bibinfo{author}{Allison, J.P.},
  \bibinfo{year}{2015}.
\newblock \bibinfo{title}{Immune checkpoint targeting in cancer therapy:
  Towards combination strategies with curative potential}.
\newblock \bibinfo{journal}{cell} \bibinfo{volume}{162},
  \bibinfo{pages}{205--214}.
\bibitem[{Sotolongo-Costa et~al.(2003)Sotolongo-Costa, Molina, Perez, Antoranz
  and Reyes}]{Costa2003}
\bibinfo{author}{Sotolongo-Costa, O.}, \bibinfo{author}{Molina, L.M.},
  \bibinfo{author}{Perez, D.R.}, \bibinfo{author}{Antoranz, J.C.},
  \bibinfo{author}{Reyes, M.C.}, \bibinfo{year}{2003}.
\newblock \bibinfo{title}{{Behavior of tumors under nonstationary therapy}}.
\newblock \bibinfo{journal}{Physica D: Nonlinear Phenomena}
  \bibinfo{volume}{178}, \bibinfo{pages}{242--253}.
\newblock \href{http://arxiv.org/abs/0203057}{\tt arXiv:0203057}.
\bibitem[{Talkington et~al.(2018)Talkington, Dantoin and
  Durrett}]{Talkington2018}
\bibinfo{author}{Talkington, A.}, \bibinfo{author}{Dantoin, C.},
  \bibinfo{author}{Durrett, R.}, \bibinfo{year}{2018}.
\newblock \bibinfo{title}{Ordinary differential equation models for adoptive
  immunotherapy}.
\newblock \bibinfo{journal}{Bull. Math. Biol.} \bibinfo{volume}{80},
  \bibinfo{pages}{1059--1083}.
\bibitem[{Wherry(2011)}]{DWherry2011}
\bibinfo{author}{Wherry, E.J.}, \bibinfo{year}{2011}.
\newblock \bibinfo{title}{T cell exhaustion}.
\newblock \bibinfo{journal}{Nature Immunology} \bibinfo{volume}{12},
  \bibinfo{pages}{492–499}.
\bibitem[{Wherry and Kurachi(2015)}]{Wherry2015}
\bibinfo{author}{Wherry, E.J.}, \bibinfo{author}{Kurachi, M.},
  \bibinfo{year}{2015}.
\newblock \bibinfo{title}{Molecular and cellular insights into {T} cell
  exhaustion}.
\newblock \bibinfo{journal}{Nature reviews. Immunology}
  \bibinfo{volume}{15(8)}, \bibinfo{pages}{486–499}.
\bibitem[{Zhang and Wang(2019)}]{Zhang2019}
\bibinfo{author}{Zhang, J.}, \bibinfo{author}{Wang, L.}, \bibinfo{year}{2019}.
\newblock \bibinfo{title}{The emerging world of {TCR-T} cell trials against
  cancer: A systematic review}.
\newblock \bibinfo{journal}{Technol. Cancer Res. Treat.} \bibinfo{volume}{18},
  \bibinfo{pages}{1--13}.

\end{thebibliography}

\end{document}